\documentclass[12pt]{article}
\pdfoutput=1

\usepackage{amsmath}
\usepackage{amssymb}
\usepackage{epsfig}
\usepackage{epstopdf}
\usepackage{latexsym}
\usepackage{color}
\numberwithin{equation}{section}
\usepackage{multirow}
\usepackage[
      colorlinks=false,
      linkcolor=darkblue,  
      urlcolor=blue,    
      filecolor=blue,     
      citecolor=red,
linktocpage=true,
      pdfstartview=FitV,
      bookmarksopen=true    
      ]{hyperref}

\begin{document}

\begin{titlepage}

\centerline
\centerline
\centerline
\bigskip
\bigskip
\centerline{\Huge \rm M5-branes and D4-branes wrapped on}
\bigskip
\centerline{\Huge \rm disk $\times$ disk and spindle $\ltimes$ disk}
\bigskip
\bigskip
\bigskip
\bigskip
\bigskip
\bigskip
\bigskip
\bigskip
\centerline{\rm Minwoo Suh}
\bigskip
\centerline{\it School of General Education, Kumoh National Institute of Technology,}
\centerline{\it Gumi, 39177, Korea}
\bigskip
\centerline{\tt minwoosuh1@gmail.com} 
\bigskip
\bigskip
\bigskip
\bigskip
\bigskip
\bigskip
\bigskip
\bigskip

\begin{abstract}
\noindent We construct and study the $AdS_3\times\text{disk}\times\text{disk}$ and $AdS_2\times\text{disk}\times\text{disk}$ solutions of $U(1)^2$-gauged supergravity in seven and six dimensions, respectively. For the construction of $AdS_3\times\text{disk}\times\text{disk}$ solutions, we employ the previously constructed consistent truncation of seven-dimensional gauged supergravity on a disk. We uplift the solutions to eleven-dimensional and massive type IIA supergravity, respectively, and study the disk geometry of the solutions. We perform flux quantizations and calculate the holographic central charge and the Bekenstein-Hawking entropy, respectively. In a similar manner, we present the $AdS_3\times\text{spindle}\ltimes\text{disk}$ and $AdS_2\times\text{spindle}\ltimes\text{disk}$ solutions.
\end{abstract}

\vskip 5cm

\flushleft {November, 2024}

\end{titlepage}

\tableofcontents

\section{Introduction}

Recently, supersymmetric $AdS$ solutions obtained from wrapping branes on orbifolds have been constructed and studied actively. In particular, two classes of orbifolds were considered. The first class is the spindle and it is topologically a sphere with two orbifold singularities at its poles, \cite{Ferrero:2020laf, Ferrero:2020twa}. The second class is the disk and it has a single orbifold singularity at its center, \cite{Bah:2021mzw, Bah:2021hei}. However, the spindle and the disk solutions share the identical local form of the solution and the solutions are obtained by different global completions of the local solution.

We start by reviewing the landscape of $AdS\times\text{disk}$ solutions. Let us consider $U(1)^4$-gauged supergravity in four dimensions which is also known as the STU model. There are four $U(1)$ gauge fields, $A^I$, $I=1,2,3,4$, in the model. For the special case, $A^1=A^2=A^3$ and $A^4=0$, we find the $AdS_2\times\text{disk}$ solution, \cite{Suh:2021hef, Couzens:2021rlk}. Also consider $U(1)^3$-gauged supergravity in five dimensions which is also known as the STU model. There are three $U(1)$ gauge fields, $A^I$, $I=1,2,3$, in the model. For the special case, $A^1=A^2$ and $A^3=0$, there is the $AdS_3\times\text{disk}$ solution, \cite{Couzens:2021tnv, Suh:2021ifj}. So far only one $AdS\times\text{disk}$ solution was discovered in four and five dimensions, respectively. See also \cite{Boisvert:2024jrl} for non-$AdS$ disk solutions.

Now we consider six-dimensional gauged supergravity coupled to a vector multiplet. There are one $U(1)$ gauge field from the gravity multiplet, $A^1$, and another $U(1)$ gauge field from the vector multiplet, $A^2$.{\footnote{Notice that the notation of the gauge fields is different from the one used in the main body of the paper. The two gauge fields in the main body are $A_1\equiv{A}^1+A^2$ and $A_2\equiv{A}^1-A^2$ by the notation here in the introduction.}} If we have $A^1\ne0$ and $A^2=0$, the theory reduces to minimal gauged supergravity in six dimensions. Unlike in four and five dimensions, there are two distinct classes of $AdS_4\times\text{disk}$ solutions in six dimensions. One is from the case of minimal gauged supergravity, $A^1\ne0$ and $A^2=0$, and we will refer to it as the $minimal$ disk solution, \cite{Suh:2021aik}, (or, equivalently, as the $pure$ disk solution.) Another one is from the case of $A^1=A^2\ne0$ and we will refer to it as the $maximal$ disk solution, (or, equivalently, as the $matter-coupled$ disk solution.). It was studied in appendix C.1 of \cite{Couzens:2022lvg} and we will further consider its flux quantization and the calculation of holographic free energy in appendix \ref{appD}. 

It is parallel in seven dimensions and there are also two classes of $AdS_5\times\text{disk}$ solutions. The maximal $AdS_5\times\text{disk}$ solution was studied in \cite{Bah:2021mzw, Bah:2021hei} and, via the AdS/CFT correspondence, \cite{Maldacena:1997re}, it was proposed to be dual to a class of Argyres-Douglas theories, \cite{Argyres:1995jj}. See \cite{Karndumri:2022wpu} and \cite{Couzens:2022yjl, Bah:2022yjf, Bomans:2023ouw, Couzens:2023kyf} for further studies. The minimal solution has not been considered in the literature. We summarize the classification of $AdS\times\text{disk}$ solutions in Table \ref{table1}.

\begin{table}[t]
\centering
\begin{tabular}{|c|c|c|c|c|} 
\hline
Dimensions & Solutions & Gauge fields & References \\

\hline \hline
$D=4$ & $AdS_2\times\text{disk}$ & $A^1=A^2=A^3$, $A^4=0$ & \cite{Suh:2021hef, Couzens:2021rlk} \\
\cline{2-2} \cline{4-4}

\hline
$D=5$ & $AdS_3\times\text{disk}$ & $A^1=A^2$, $A^3=0$ & \cite{Couzens:2021tnv, Suh:2021ifj} \\
\cline{2-2} \cline{4-4}
                 
\hline
$D=6$ & $AdS_4\times\text{disk}$ & $A^1\ne0$, $A^2=0$ & \cite{Suh:2021aik} \\
\cline{3-4}
                 &  & $A^1=A^2$ & \cite{Couzens:2022lvg}, [Here] \\
\cline{2-4}

\hline
$D=7$ & $AdS_5\times\text{disk}$ & $A^1\ne0$, $A^2=0$ & \\
\cline{3-4}
                 &  & $A^1=A^2$ & \cite{Bah:2021mzw, Bah:2021hei}  \\
\hline
\end{tabular}
\caption{Classification of $AdS\times\text{disk}$ solutions.}
\label{table1}
\end{table}

Furthermore, one can consider $AdS\times\text{disk}\times\text{disk}$ solutions. Recently, consistent truncation of seven-dimensional gauged supergravity on the maximal $AdS_5\times\text{disk}$ solution to five-dimensional gauged supergravity was performed in \cite{Bomans:2023ouw}. Employing this result, one can uplift the $AdS_3\times\text{disk}$ solution, \cite{Couzens:2021tnv, Suh:2021ifj}, to seven dimensions and obtain the $AdS_3\times\text{disk}\times\text{disk}$ solution.

In this paper, by employing the consistent truncation of \cite{Bomans:2023ouw}, we construct the $AdS_3\times\text{disk}\times\text{disk}$ solutions and study the disk geometry. We uplift the solutions to eleven-dimensional supergravity and perform flux quantization and calculate the holographic central charge. As the $AdS_5\times\text{disk}$ solution is dual to a class of 4d Argyres-Douglas theories, \cite{Bah:2021mzw, Bah:2021hei}, the $AdS_3\times\text{disk}\times\text{disk}$ solutions can be thought to be dual to 2d SCFTs from 4d Argyres-Douglas theories compactified on a disk.

Furthermore, inspired by the $AdS_3\times\text{disk}\times\text{disk}$ solutions, we construct and study the $AdS_2\times\text{disk}\times\text{disk}$ solution of six-dimensional $U(1)^2$-gauged supergravity. We construct the $AdS_2\times\text{disk}\times\text{disk}$ solution with two non-trivial distinct $U(1)$ gauge fields by trial and error. We study the solution and find usual properties of the disk geometry, $e.g.$, the Euler characteristic of the disk. We further uplift the solution to massive type IIA supergravity and find the monopole and the smeared D4-D8-branes from the structure of the solution. Furthermore, we perform the flux quantization and calculate the Bekenstein-Hawking entropy of the presumed black hole.

In the same manner, we also construct $AdS_3\times\text{spindle}\ltimes\text{disk}$ and $AdS_2\times\text{spindle}\ltimes\text{disk}$ solutions. We refrain from detailed analysis of the solutions, as they are parallel to the ones for $AdS_{2,3}\times\text{disk}\times\text{disk}$ solutions. However, we calculate the holographic central charge of the $AdS_3\times\text{spindle}\ltimes\text{disk}$ solutions. In table \ref{table2} we summarize the so far studied classification of $AdS_{2,3}\times{M}_4$ solutions from $U(1)^2$-gauged supergravity in six and seven dimensions.

In section \ref{sec2}, we construct and study the $AdS_3\times\text{disk}\times\text{disk}$ solutions. In section \ref{sec3}, we present the $AdS_3\times\text{spindle}\ltimes\text{disk}$ solutions. In section \ref{sec4}, we construct and study the $AdS_2\times\text{disk}\times\text{disk}$ solutions. In section \ref{sec5}, we present the $AdS_2\times\text{spindle}\ltimes\text{disk}$ solutions. In section \ref{sec6} we conclude with open questions. In appendix \ref{appA} the equations of motion of seven- and six-dimensional $U(1)^2$-gauged supergravity are presented. In appendix \ref{appB} we review the $AdS_3\times\text{spindle}$ and $AdS_3\times\text{disk}$solutions. In appendix \ref{appC} we review the $AdS_2\times\text{spindle}$ and $AdS_2\times\text{disk}$ solutions. In appendix \ref{appD} we review the maximal $AdS_4\times\text{disk}$ solutions.

\begin{table}[t]
\centering
\begin{tabular}{|c|c|} 
\hline
Solutions & References \\

\hline \hline
$AdS_3\times\text{Riemann}\times\text{Riemann}$ & \cite{Gauntlett:2001jj} \\
\cline{2-2}

\hline
$AdS_2\times\text{Riemann}\times\text{Riemann}$ & \cite{Suh:2018tul, Hosseini:2018usu, Suh:2018szn, Kim:2019fsg} \\
\cline{2-2}
                 
\hline \hline
$AdS_3\times\text{spindle}\times\text{Riemann}$ & \cite{Boido:2021szx, Suh:2022olh} \\
\cline{2-2}

\hline
$AdS_2\times\text{spindle}\times\text{Riemann}$ & \cite{Giri:2021xta, Faedo:2021nub, Suh:2022olh} \\
\cline{2-2}

\hline \hline
$AdS_3\times\text{spindle}\ltimes\text{spindle}$ & \cite{Cheung:2022ilc} \\
\cline{2-2}

\hline
$AdS_2\times\text{spindle}\ltimes\text{spindle}$ & \cite{Couzens:2022lvg, Faedo:2022rqx} \\
\cline{2-2}

\hline \hline
$AdS_3\times\text{Riemann}\ltimes\text{spindle}$ & \cite{Cheung:2022ilc} \\
\cline{2-2}

\hline
$AdS_2\times\text{Riemann}\ltimes\text{spindle}$ & \cite{Couzens:2022lvg, Faedo:2022rqx} \\
\cline{2-2}

\hline \hline
$AdS_3\times\text{quadrilateral}$ & \cite{Faedo:2024upq} \\
\cline{2-2}

\hline
$AdS_2\times\text{quadrilateral}$ & \cite{Faedo:2024upq} \\
\cline{2-2}

\hline \hline
$AdS_3\times\text{disk}\times\text{Riemann}$ & \cite{Suh:2021ifj} \\
\cline{2-2}

\hline
$AdS_2\times\text{disk}\times\text{Riemann}$ &  \\
\cline{2-2}

\hline \hline
$AdS_3\times\text{disk}\times\text{disk}$ & \cite{Bomans:2023ouw}, [Here] \\
\cline{2-2}

\hline
$AdS_2\times\text{disk}\times\text{disk}$ & [Here] \\
\cline{2-2}

\hline \hline
$AdS_3\times\text{spindle}\ltimes\text{disk}$ & \cite{Bomans:2023ouw}, [Here] \\
\cline{2-2}

\hline
$AdS_2\times\text{spindle}\ltimes\text{disk}$ & [Here] \\
\cline{2-2}
\hline

\end{tabular}
\caption{Classification of $AdS_{2,3}\times{M}_4$ solutions from $U(1)^2$-gauged supergravity in six and seven dimensions. The Riemann denotes a Riemann surface with genus, $\mathfrak{g}$.}
\label{table2}
\end{table}

\section{M5-branes wrapped on disk $\times$ disk} \label{sec2}

\subsection{$U(1)^2$-gauged supergravity in seven dimensions}

We review $U(1)^2$-gauged supergravity in seven dimensions, \cite{Liu:1999ai}, in the conventions of \cite{Cheung:2022ilc}. The bosonic field content is consist of the metric, two $U(1)$ gauge fields, $A^{12}$, $A^{34}$, a three-form field, $S^5$, and two scalar fields, $\lambda_1$, $\lambda_2$. The Lagrangian is given by
\begin{align} \label{sevenlag}
\mathcal{L}\,=\,&\left(R-V\right)\text{vol}_7-6*d\lambda_1\wedge\,d\lambda_1-6*d\lambda_2\wedge\,d\lambda_2-8*d\lambda_1\wedge\,d\lambda_2 \notag \\
&-\frac{1}{2}e^{-4\lambda_1}*F^{12}\wedge\,F^{12}-\frac{1}{2}e^{-4\lambda_2}*F^{34}\wedge\,F^{34}-\frac{1}{2}e^{-4\lambda_1-4\lambda_2}*S^5\wedge\,S^5 \notag \\
&+\frac{1}{2g}S^5\wedge\,dS^5-\frac{1}{g}S^5\wedge\,F^{12}\wedge\,F^{34}+\frac{1}{2g}A^{12}\wedge\,F^{12}\wedge\,F^{34}\wedge\,F^{34}\,,
\end{align}
where $F^{12}=dA^{12}$, $F^{34}=dA^{34}$ and the scalar potential is
\begin{equation}
V\,=\,g^2\left[\frac{1}{2}e^{-8\left(\lambda_1+\lambda_2\right)}-4e^{2\left(\lambda_1+\lambda_2\right)}-2e^{-2\left(2\lambda_1+\lambda_2\right)}-2e^{-2\left(\lambda_1+2\lambda_2\right)}\right]\,.
\end{equation}
The equations of motion are presented in appendix \ref{appA1}.

We introduce a parametrization of the scalar and gauge fields,
\begin{align}
&X_1\,=\,e^{2\lambda_1}\,, \qquad X_2\,=\,e^{2\lambda_2}\,, \notag \\
&A_1\,=\,A^{12}\,, \qquad A_2\,=\,A^{34}\,.
\end{align}
and set the gauge coupling constant, $g=1$, for the rest of the work.

\subsection{The $AdS_3\times\text{disk}\times\text{disk}$ solutions}

The consistent truncation of seven-dimensional maximal gauged supergravity, \cite{Pernici:1984xx}, on a disk was performed in \cite{Bomans:2023ouw}. By uplifting the $AdS_3\times\text{disk}$ solutions, \cite{Couzens:2021tnv, Suh:2021ifj}, which we review in appendix \ref{appB}, the $AdS_3\times\text{disk}\times\text{disk}$ solutions can be readily obtained.{\footnote{In \cite{Bomans:2023ouw} it appears that some additional numerical factors are required for the uplift ansatz: $ds_5^2\rightarrow4ds_5^2$ in (4.19) and $A^{34}=2\mathcal{A}^3$, $A^{45}=2\mathcal{A}^1$, $A^{53}=2\mathcal{A}^2$ in (4.22) of \cite{Bomans:2023ouw}. This could be due to the conventional difference of the uplift ansatz and the solution to be uplifted in \cite{Bomans:2023ouw}.}}{\footnote{We also fix a typographical error in the last equation in (4.23) of \cite{Bomans:2023ouw}: $w^2\rightarrow{w}^3$.}}

The local form of the $AdS_3\times\text{disk}\times\text{disk}$ solution is
\begin{align} \label{ads3one}
ds_7^2\,=&\,\frac{y}{XX_1^{1/3}}\left[f^{1/3}\left(ds_{AdS_3}^2+\frac{1}{4p}dx^2+\frac{p}{f}d\psi^2\right)+\frac{yX}{4F}dy^2+\frac{FX^4}{F+y^3X^3}dz^2\right]\,, \notag \\
X_1\,=&\,\left[\frac{y^4}{4\left(F+y^3X^3\right)}\right]^{3/5}\,, \qquad X_2\,=\,\left[\frac{y^4}{4\left(F+y^3X^3\right)}\right]^{-2/5}\,, \notag \\
A_1\,=&\,-\frac{y^4X^3}{2\left(F+y^3X^3\right)}dz\,, \qquad A_2\,=\,\frac{2x}{x-s_1}d\psi\,, \notag \\
S^5\,=&\,\sqrt{2}s_1y\,\text{vol}_{AdS_3}-\frac{yF}{\sqrt{2}\left(F+y^3X^3\right)}\frac{s_1}{\left(x-s_1\right)^2}dx\wedge{d}\psi\wedge{d}z\,,
\end{align}
where $q_1$ and $s_1$ are constant parameters. We have
\begin{equation}
F(y)\,=\,\frac{1}{4}\left(y^2+q_1\right)y^2-y^3\,.
\end{equation}
and
\begin{equation}
f(x)\,=\,\left(x-s_1\right)^2x\,, \qquad p(x)\,=\,f(x)-x^2\,, \qquad X(x)\,=\,\frac{f(x)^{1/3}}{x-s_1}\,.
\end{equation}
Thus the solution has a non-trivial scalar field and two $U(1)$ gauge fields, effectively.

The solution can also be given in the form of
\begin{align}
ds_7^2\,=&\,\frac{y^{3/5}\left(4\left(h+yX^3\right)\right)^{1/5}}{X}\left[f^{1/3}\left(ds_{AdS_3}^2+\frac{1}{4p}dx^2+\frac{p}{f}d\psi^2\right)+\frac{X}{4yh}dy^2+\frac{hX^4}{h+yX^3}dz^2\right]\,, \notag \\
X_1\,=&\,\left[\frac{y^2}{4\left(h+yX^3\right)}\right]^{3/5}\,, \qquad X_2\,=\,\left[\frac{y^2}{4(h+yX^3)}\right]^{-2/5}\,, \notag \\
A_1\,=&\,-\frac{y^2X^3}{2(h+yX^3)}dz\,, \qquad A_2\,=\,\frac{2x}{x-s_1}d\psi\,, \notag \\
S^5\,=&\,\sqrt{2}s_1y\,\text{vol}_{AdS_3}-\frac{yh}{\sqrt{2}\left(h+yX^3\right)}\frac{s_1}{\left(x-s_1\right)^2}dx\wedge{d}\psi\wedge{d}z\,,
\end{align}
where we define
\begin{equation}
h(y)\,=\,\frac{1}{4}\left(y^2+q_1\right)-y\,.
\end{equation}

Now we consider the global completion of the solution, $i.e.$, find the range of the solution where the metric functions are positive definite and the fields are real. We find such solutions when
\begin{equation}
0<y<y_1\,,
\end{equation}
where we find
\begin{align} \label{yone}
y_1\,=2-\sqrt{4-q_1}\,,
\end{align}
and $y_1$ is a solution of $h(y)=0$.

Near $y\rightarrow0$, the warp factor vanishes and it is a curvature singularity of the metric,
\begin{equation}
ds_7^2\,\approx\,\frac{q_1^{1/5}y^{3/5}}{X}\left[f^{1/3}\left(ds_{AdS_3}^2+\frac{1}{4p}dx^2+\frac{p}{f}d\psi^2\right)+\frac{X}{q_1y}dy^2+X^4dz^2\right]\,.
\end{equation}

Approaching $y\rightarrow{y}_1$, the metric becomes to be
\begin{equation}
ds_7^2\,\approx\,\frac{4^{1/5}y_1^{4/5}}{X^{2/5}}\left[f^{1/3}\left(ds_{AdS_3}^2+\frac{1}{4p}dx^2+\frac{p}{f}d\psi^2\right)+\frac{X\left[d\rho^2+\mathcal{E}^2(q_1)\rho^2dz^2\right]}{y_1\left(-h'(y_1)\right)}\right]\,,
\end{equation}
where we introduced a new coordinate, $\rho^2\,=\,y_1-y$. The function, $\mathcal{E}(q_1)$, is given by
\begin{equation} \label{eq1}
\mathcal{E}(q_1)\,=\,\sqrt{1-\frac{q_1}{4}}\,,
\end{equation}
where we evaluated it at $x=x_1$ and $y=y_1$ where both disks are at their orbifold singularities. Then, the $y-z$ surface is locally an $\mathbb{R}^2/\mathbb{Z}_\ell$ orbifold if we set
\begin{equation} \label{mathcalE}
\mathcal{E}(q_1)\,=\,\frac{1}{\mathcal{C}\ell}\,=\,\frac{2\pi}{\Delta{z}\ell}\,,
\end{equation}
where $\Delta{z}$ is the period of the coordinate, $z$, and $\ell=1,2,3,\ldots$.{\footnote{The range of the $z$ coordinate, $\Delta{z}$, is commonly employed in the spindle literature and the parameter, $\mathcal{C}$, is in the disk literature and they are functions of $q_1$. Here we present their equivalence up to the factor of $2\pi$.}} Thus the metric on the surface, $\Sigma(y,z)$, has a topology of disk with the origin at $y=y_1$ and the boundary at $y=0$.

We calculate the Euler characteristic of the surface, $\Sigma(y,z)$,
\begin{align}
\chi\left(\Sigma\right)\,=&\,\frac{1}{4\pi}\int_\Sigma{R}_\Sigma\text{vol}_\Sigma\,=\,\frac{2\pi}{4\pi}\frac{4y_1^{1/2}\left(y_1^2-q_1\right)X(x_1)^{9/2}}{\left(y_1^2+q_1-4y_1\left(1-X(x_1)^3\right)\right)^{3/2}}\frac{\Delta{z}}{2\pi} \notag \\
=&\,\mathcal{E}\left(q_1\right)\frac{\Delta{z}}{2\pi}\,=\,\mathcal{C}\mathcal{E}\left(q_1\right)\,=\,\frac{1}{\ell}\,,
\end{align}
which is a natural result for a disk.

\subsection{Uplift to eleven-dimensional supergravity}

Employing the uplift formula in \cite{Cvetic:2000ah} presented in \cite{Cheung:2022ilc} with the choice of
\begin{equation}
w_0\,=\,\sin\xi\,, \qquad w_1\,=\,\cos\xi\cos\theta\,, \qquad w_2\,=\,\cos\xi\sin\theta\,,
\end{equation}
and $\xi,\theta\in\left[0,\pi/2\right]$, we uplift the $AdS_3\times\text{disk}\times\text{disk}$ solutions to eleven-dimensional supergravity, \cite{Cremmer:1978km}. The metric is given by
\begin{align}
ds_{11}^2\,=\,\frac{\tilde{\Delta}^{1/3}y^{1/3}}{X}&\left[f^{1/3}\left(ds_{AdS_3}^2+\frac{1}{4p}dx^2+\frac{p}{f}d\psi^2\right)+\frac{X}{4yh}dy^2+\frac{hX^4}{h+yX^3}dz^2\right. \notag \\
&+\left.\frac{X}{y}d\xi^2+\frac{X}{\tilde{\Delta}y}\Big(4\left(h+yX^3\right)\sin^2\xi{D}\phi_1^2+y^2\cos^2\xi\left(d\theta^2+\sin^2\theta{D}\phi_2^2\right)\Big)\right]\,,
\end{align}
where $D\phi_i\equiv{d}\phi_i-A_i$ and $\phi_1,\phi_2\in\left[0,2\pi\right]$. The four-form flux is given by
\begin{align}
F_{(4)}\,=\,&\frac{4\left(h+yX^3\right)\left[2\left(2h+y\left(y+2X^3\right)\right)\cos^2\xi+3y^2\sin^2\xi\right]}{\left[4\left(h+yX^3\right)\cos^2\xi+y^2\sin^2\xi\right]^2} \notag \\
&\times\sin\xi\cos^2\xi\sin\theta{d}\xi\wedge{d}\theta\wedge{D}\phi_1\wedge{D}\phi_2 \notag \\
+&F_{1yz}\frac{4\left(h+yX^3\right)}{4\left(h+yX^3\right)\cos^2\xi+y^2\sin^2\xi}\cos^2\xi\sin\theta{d}y\wedge{d}z\wedge{d}\theta\wedge{D}\phi_2 \notag \\
+&F_{2x\psi}\frac{y^2}{4\left(h+yX^3\right)\cos^2\xi+y^2\sin^2\xi}\sin^2\xi\cos\xi\sin\theta{d}x\wedge{d}\psi\wedge{d}\theta\wedge{D}\phi_1 \notag \\
+&F_{2x\psi}\sin\xi\cos\theta{d}x\wedge{d}\psi\wedge{d}\xi\wedge{D}\phi_1 \notag \\
+&\ldots\,
\end{align}
where $F_1$ and $F_2$ are the field strengths of seven-dimensional $U(1)$ gauge fields. We have only presented the terms required for the flux quantization.

\subsection{Uplifted metric}

\begin{figure}[t] 
\begin{center} 
\includegraphics[width=4.5in]{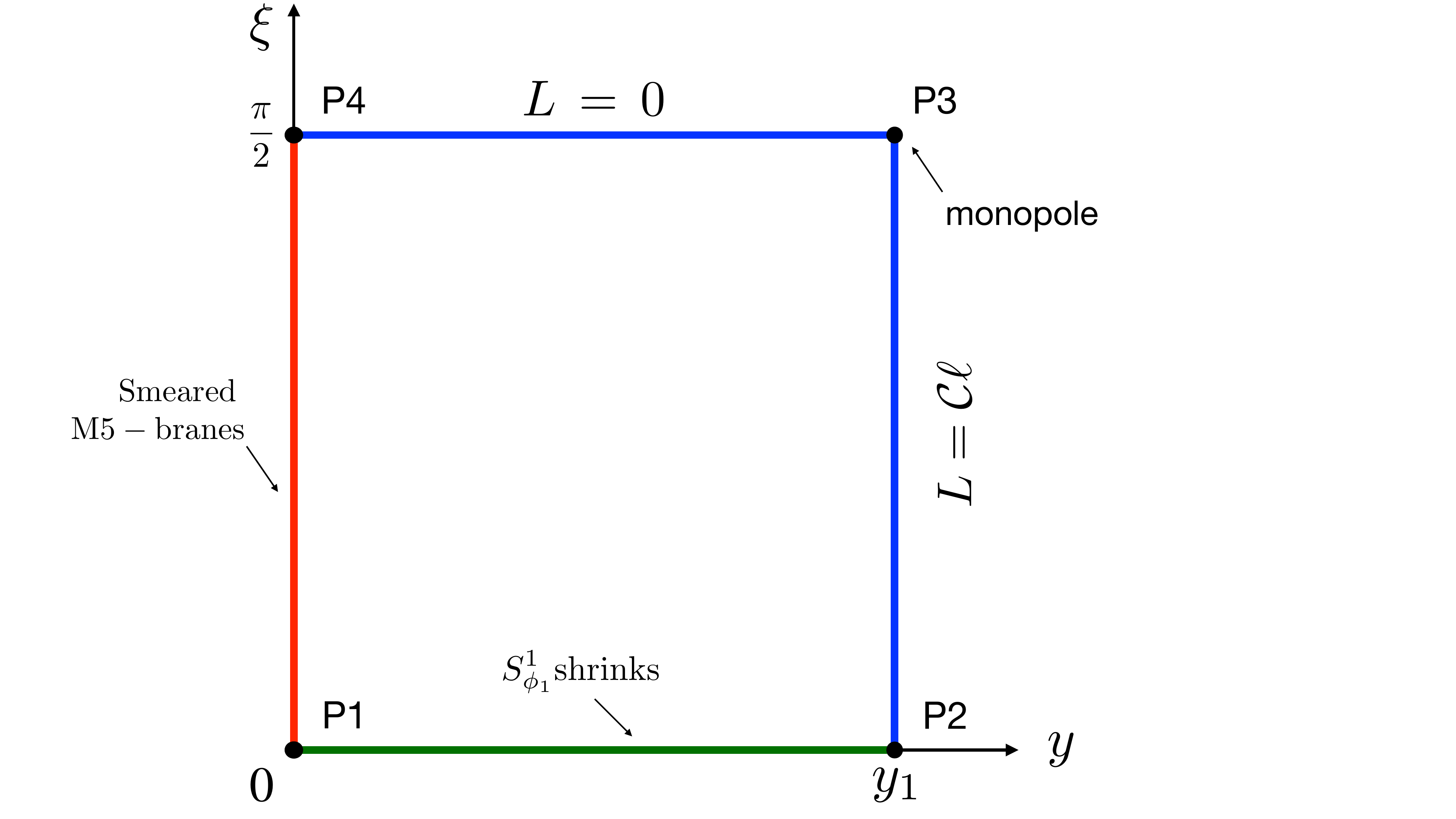} 
\caption{{\it The two-dimensional base space, $B_2$, spanned by $y$ and $\xi$.}} \label{fig1}
\end{center}
\end{figure}

The six-dimensional internal space of the uplifted metric is an $S_z^1\,\times\,S^3$ fibration over the 2d base space, $B_2$, of $(y,\xi)$. The 2d base space is a rectangle of $(y,\xi)$ over $[0,y_1)\,\times\left[0,\frac{\pi}{2}\right]$. See Figure \ref{fig1}. We explain the geometry of the internal space by three regions of the 2d base space, $B_2$.

\begin{itemize}
\item Region I: The side of $\mathsf{P}_1\mathsf{P}_2$.
\item Region II: The sides of $\mathsf{P}_2\mathsf{P}_3$ and $\mathsf{P}_3\mathsf{P}_4$.
\item Region III: The side of $\mathsf{P}_1\mathsf{P}_4$.
\end{itemize}

\noindent {\bf Region I:} On the side of $\xi\,=\,0$, the $\phi_1$ circle, $S^1_{\phi_1}$, shrinks.

\bigskip

\noindent {\bf Region II: Monopole} In order to observe the property of the 2d base, \eqref{2db3}, we fix the gauge of the six-dimensional gauge field to be
\begin{equation}
A_1\,=\,-\frac{y^2X^3}{2\left(h+yX^3\right)}dz+1\,.
\end{equation}
We break $D\phi_1^2=\left(d\phi_1-A_1\right)^2$ and complete the square of $dz$ to obtain the metric of
\begin{align}
ds_{11}^2\,=&\,\frac{\tilde{\Delta}^{1/3}y^{1/3}}{X}\left[f^{1/3}\left(ds_{AdS_3}^2+\frac{1}{4p}dx^2+\frac{p}{f}d\psi^2\right)+\frac{X}{4yh}dy^2\right. \notag \\
& \qquad +\frac{X}{y}d\xi^2+\frac{X}{\tilde{\Delta}y}y^2\cos^2\xi\left(d\theta^2+\sin^2\theta{D}\phi_2^2\right)\,, \\
& \qquad +R_z^2\left(dz+Ld\phi_1\right)^2+R_{\phi_1}^2d\phi_1^2\Big]\,,
\end{align}
where we define
\begin{align}
R_z^2\,=&\,\frac{XY}{\tilde{\Delta}y}\,, \notag \\
R_{\phi_1}^2\,=&\,\frac{4hX^4\sin^2\xi}{Y}\,, \notag \\
L\,=&\,\frac{2\left(2h-y\left(y-2\right)X^3\right)\sin^2\xi}{Y}\,,
\end{align}
with
\begin{align}
Y\,\equiv\,&4yhX^3\cos^2\xi+\left(4h+y\left(y-2\right)^2X^3\right)\sin^2\xi\,.
\end{align}

The function, $L(y,\xi)$, is piecewise constant along the sides of $y=y_1$ and $\xi=\frac{\pi}{2}$ of the 2d base, $B_2$, 
\begin{equation} \label{2db3}
L\left(y,\frac{\pi}{2}\right)\,=\,0\,, \qquad L(y_1,\xi)\,=\,\frac{1}{\mathcal{E}(q_1)}\,=\,\mathcal{C}\ell\,=\,\frac{\Delta{z}}{2\pi}\ell\,.
\end{equation}
The jump at the corner, $(y,\xi)=\left(y_1,\frac{\pi}{2}\right)$, indicates a monopole source for the $D\phi_1$ fibration.

\bigskip

\noindent {\bf Region III: Smeared M5-branes} In the limit, $y\rightarrow0$, the metric asymptotes to
\begin{align}
ds_{11}^2\,\approx&\,\frac{q_1^{1/3}y^{1/3}\cos^{1/3}\xi}{X}\left[f^{1/3}\left(ds_{AdS_3}^2+\frac{1}{4p}dx^2+\frac{p}{f}d\psi^2\right)+X^4dz^2\right. \notag \\
&+\left.\frac{X}{q_1y}\Big(q_1\left(d\xi^2+\tan^2\xi{D}\phi_1^2\right)+dy^2+y^2\left(d\theta^2+\sin^2\theta{D}\phi_2^2\right)\Big)\right]\,.
\end{align}
The metric implies the smeared M5-brane sources. The M5-branes are 
\begin{itemize}
\item extended along the $AdS_3$, $x$, $\psi$, and $z$ directions;
\item localized at the origin of the $\mathbb{R}^3$ parametrized by $S^2$ and $y$, $ds^2_{\mathbb{R}^3}=dy^2+y^2ds^2_{S^2}$;
\item smeared along the $\xi$ and $\phi_1$ directions. 
\end{itemize}

\subsection{Flux quantization}

The integral of four-form flux through any four-cycle is an integer,
\begin{equation}
\frac{1}{\left(2\pi\ell_p\right)^3}\int_{M_4}F_{(4)}\,\in\,\mathbb{Z}\,,
\end{equation}
where $\ell_p$ is the Planck length. First, from the flux through the four-cylce of $\xi\theta\phi_1\phi_2$, we obtain
\begin{equation}
\frac{1}{\left(2\pi\ell_p\right)^3}\int_{\xi\theta\phi_1\phi_2}{F}_{(4)}\,=\,\frac{1}{2\pi\ell_p^3}\,\equiv\,N\,,
\end{equation}
where $N\in\mathbb{N}$ is the number of M5-branes wrapping the disk$\times$disk.

Second, from the flux through the four cycle of $yz\theta\phi_2$ including the disk, $\Sigma(y,z)$, at $\xi=0$, we find
\begin{equation} \label{qKd3}
\frac{1}{\left(2\pi\ell_p\right)^3}\int_{yz\theta\phi_2}{F}_{(4)}\,=\,\frac{1}{\left(2\pi\ell_p\right)^3}\frac{\left(2\pi\right)^2}{\ell}\frac{1-\sqrt{1-\frac{q_1}{4}}}{\sqrt{1-\frac{q_1}{4}}}\,=\,\frac{N}{\ell}\frac{1-\sqrt{1-\frac{q_1}{4}}}{\sqrt{1-\frac{q_1}{4}}}\,\equiv\,K\,,
\end{equation}
where $K\in\mathbb{Z}$ is another integer. We can solve for $q_1$ and $\mathcal{C}$ in terms of the quantum numbers, $N$, $K$, and $\ell$,
\begin{equation}
\sqrt{1-\frac{q_1}{4}}\,=\,\frac{N}{N+K\ell}\,, \qquad \mathcal{C}\,=\,\frac{N+K\ell}{N\ell}\,.
\end{equation}

Third, from the flux through the four cycle of $x\psi\xi\phi_1$ including the disk, $\Sigma(x,\psi)$, at $\theta=0$, we find
\begin{equation} \label{qK3d3}
\frac{1}{\left(2\pi\ell_p\right)^3}\int_{x\psi\xi\phi_1}{F}_{(4)}\,=\,\frac{1}{\left(2\pi\ell_p\right)^3}\frac{8\pi^2}{k}\frac{1+2s_1-\sqrt{1+4s_1}}{1+4s_1-\sqrt{1+4s_1}}\,=\,\frac{2N}{k}\frac{1+2s_1-\sqrt{1+4s_1}}{1+4s_1-\sqrt{1+4s_1}}\,\equiv\,M\,,
\end{equation}
where $M\in\mathbb{Z}$ is another integer and the orbifold number, $k$, is introduced in appendix \ref{appB}. The flux quantiztions on the four cycles, $yz\theta\phi_2$ in \eqref{qKd3} and $x\psi\xi\phi_1$ in \eqref{qK3d3}, effectively reduce to flux quantizations of seven-dimensional field strengths, $F_1$ and $F_2$, respectively. We can solve for $s_1$ and $\mathcal{D}$ in terms of the quantum numbers, $N$, $M$, and $k$,
\begin{equation}
\sqrt{1+4s_1}\,=\,\frac{N}{N+Mk}\,, \qquad \mathcal{D}\,=\,\frac{N+Mk}{Nk}\,.
\end{equation}

\subsection{Holographic central charge}

For the metric,
\begin{equation}
ds_{11}^2\,=\,e^{2\mathcal{A}}\left(ds_{AdS_3}^2+ds_{M_8}^2\right)\,,
\end{equation}
the holographic central charge is given by, $e.g.$, \cite{Boido:2021szx},
\begin{equation}
c\,=\,\frac{1}{G_N^{(11)}}\int_{M_8}e^{9\mathcal{A}}\text{vol}_{M_8}\,,
\end{equation}
where $G_N^{(11)}$ is the eleven-dimensional Newton's gravitational constant. From above, we obtain the holographic central charge,
\begin{equation} \label{holcen1}
c\,=\,\frac{16\pi}{\left(2\pi\right)^8\ell_p^9}\frac{\pi^2}{3}x_1\Delta\psi\,y_1^2\Delta{z}\,=\,\frac{N^2K^2\ell}{12\left(N+K\ell\right)}\frac{8M^2k}{N\left(N+Mk\right)}\,,
\end{equation}
where the first and second factors are from the disks, $\Sigma(y,z)$ and $\Sigma(x,\psi)$, respectively. The holographic central charge can also be written by
\begin{equation}
c\,=\,\frac{8M^2k}{N\left(N+Mk\right)}a_{4d}\,,
\end{equation}
where $a_{4d}$ is the holographic central charge of $AdS_5\times\text{disk}$ solutions, \cite{Bah:2021mzw, Bah:2021hei}. Note that if $N\sim{K}\sim{M}$, it scales as $N^3$, as the theories originating from the compactifications of 6d (2,0) theories.

\section{M5-branes wrapped on spindle $\ltimes$ disk} \label{sec3}

The local form of the $AdS_3\times\text{spindle}\ltimes\text{disk}$ solution can be easily obtained from \eqref{ads3one} by $dz\rightarrow{D}z$ and recovering $s_2$ as it is in \eqref{d3one}. As the changes do not affect most of the analysis on the $AdS_3\times\text{disk}\times\text{disk}$ solution in the previous section, we will be brief. The local form of the $AdS_3\times\text{spindle}\ltimes\text{disk}$ solution is
\begin{align}
ds_7^2\,=&\,\frac{y}{XX_1^{1/3}}\left[f^{1/3}\left(ds_{AdS_3}^2+\frac{1}{4p}dx^2+\frac{p}{f}d\psi^2\right)+\frac{yX}{4F}dy^2+\frac{FX^4}{F+y^3X^3}Dz^2\right]\,, \notag \\
X_1\,=&\,\left[\frac{y^4}{4\left(F+y^3X^3\right)}\right]^{3/5}\,, \qquad X_2\,=\,\left[\frac{y^4}{4\left(F+y^3X^3\right)}\right]^{-2/5}\,, \notag \\
A_1\,=&\,-\frac{y^4X^3}{2\left(F+y^3X^3\right)}dz\,, \qquad A_2\,=\,\frac{2x}{x-s_1}d\psi\,, \notag \\
S^5\,=&\,\sqrt{2}s_1y\,\text{vol}_{AdS_3}-\frac{yF}{\sqrt{2}\left(F+y^3X^3\right)}\frac{s_1}{\left(x-s_1\right)^2}dx\wedge{d}\psi\wedge{D}z\,,
\end{align}
where $q_1$, $s_1$, and $s_2$ are constant parameters. We have
\begin{equation}
F(y)\,=\,\frac{1}{4}\left(y^2+q_1\right)y^2-y^3\,.
\end{equation}
and
\begin{equation}
f(x)\,=\,\left(x-s_1\right)^2\left(x-s_2\right)\,, \qquad p(x)\,=\,f(x)-x^2\,, \qquad X(x)\,=\,\frac{f(x)^{1/3}}{x-s_1}\,,
\end{equation}
with
\begin{equation}
Dz\,=\,dz+\frac{1}{2}A_2\,.
\end{equation}
Thus the solution has a non-trivial scalar field and two $U(1)$ gauge fields, effectively. The spindle can be either in the twist class or in the anti-twist class depending on the choice of the parameters, \cite{Ferrero:2021etw}.

The solution can also be given in the form of
\begin{align}
ds_7^2\,=&\,\frac{y^{3/5}\left(4\left(h+yX^3\right)\right)^{1/5}}{X}\left[f^{1/3}\left(ds_{AdS_3}^2+\frac{1}{4p}dx^2+\frac{p}{f}d\psi^2\right)+\frac{X}{4yh}dy^2+\frac{hX^4}{h+yX^3}Dz^2\right]\,, \notag \\
X_1\,=&\,\left[\frac{y^2}{4\left(h+yX^3\right)}\right]^{3/5}\,, \qquad X_2\,=\,\left[\frac{y^2}{4(h+yX^3)}\right]^{-2/5}\,, \notag \\
A_1\,=&\,-\frac{y^2X^3}{2(h+yX^3)}dz\,, \qquad A_2\,=\,\frac{2x}{x-s_1}d\psi\,, \notag \\
S^5\,=&\,\sqrt{2}s_1y\,\text{vol}_{AdS_3}-\frac{yh}{\sqrt{2}\left(h+yX^3\right)}\frac{s_1}{\left(x-s_1\right)^2}dx\wedge{d}\psi\wedge{D}z\,,
\end{align}
where we define
\begin{equation}
h(y)\,=\,\frac{1}{4}\left(y^2+q_1\right)-y\,.
\end{equation}
The solution reduces back to the local form of $AdS_3\times\text{disk}\times\text{disk}$ solution in the previous section by
\begin{equation}
Dz\,\rightarrow{d}z\,, \qquad s_1\,\rightarrow\,0\,.
\end{equation}

Similar to \eqref{holcen1}, we find the holographic central charge to be
\begin{align}
c\,=&\,\frac{16\pi}{\left(2\pi\right)^8\ell_p^9}\frac{\pi^2}{3}\left(x_2-x_1\right)\Delta\psi\,y_1^2\Delta{z} \notag \\
=&\,\frac{N^2K^2\ell}{12\left(N+K\ell\right)}\frac{32q^2\left(n_--n_+-2q\right)}{n_-n_+\left[n_-\left(n_++2q\right)-q\left(2n_++3q\right)\right]}\,,
\end{align}
where the first and second factors are from the disk, $\Sigma(y,z)$ and the spindle, $\Sigma(x,\psi)$, respectively. We employed the expression, \eqref{x2x1dphi}, from \cite{Boido:2021szx}. Note that if $N\sim{K}$, it scales as $N^3$, as the theories originating from the compactifications of 6d (2,0) theories. The holographic central charge can also be written by
\begin{equation}
c\,=\,\frac{32q^2\left(n_--n_+-2q\right)}{n_-n_+\left[n_-\left(n_++2q\right)-q\left(2n_++3q\right)\right]}a_{4d}\,,
\end{equation}
where $a_{4d}$ is the holographic central charge of $AdS_5\times\text{disk}$ solutions, \cite{Bah:2021mzw, Bah:2021hei}.

\section{D4-branes wrapped on disk $\times$ disk} \label{sec4}

\subsection{$U(1)^2$-gauged supergravity in six dimensions}

We review $F(4)$ gauged supergravity, \cite{Romans:1985tw}, coupled to a vector multiplet in six dimensions, \cite{Andrianopoli:2001rs, Karndumri:2015eta}, in the conventions of \cite{Faedo:2021nub}. The bosonic field content is consist of the metric, two $U(1)$ gauge fields, $A_i$, a two-form field, $B$, and two scalar fields, $\varphi_i$, where $i\,=\,1,2$. We introduce a parametrization of the scalar fields,
\begin{equation}
X_i\,=\,e^{-\frac{1}{2}\vec{a}_i\cdot\vec{\varphi}}\,, \qquad \vec{a}_1\,=\,\left(2^{1/2},2^{-1/2}\right)\,, \qquad \vec{a}_2\,=\,\left(-2^{1/2},2^{-1/2}\right)\,,
\end{equation}
with
\begin{equation}
X_0\,=\,\left(X_1X_2\right)^{-3/2}\,.
\end{equation}
The field strengths of the gauge fields and two-form field are, respectively,
\begin{equation}
F_i\,=\,dA_i\,, \qquad H=dB\,.
\end{equation}
The action is given by
\begin{align} \label{actionaction}
S\,&=\,\frac{1}{16\pi{G}_N^{(6)}}\int{d}^6x\sqrt{-g}\left[R-V-\frac{1}{2}|d\vec{\varphi}|^2-\frac{1}{2}\sum_{i=1}^2X_i^{-2}|F_i|^2-\frac{1}{8}\left(X_1X_2\right)^2|H|^2\right. \notag \\
&\left.-\frac{m^2}{4}\left(X_1X_2\right)^{-1}|B|^2-\frac{1}{16}\frac{\varepsilon^{\mu\nu\rho\sigma\tau\lambda}}{\sqrt{-g}}B_{\mu\nu}\left(F_{1\rho\sigma}F_{2\tau\lambda}+\frac{m^2}{12}B_{\rho\sigma}B_{\tau\lambda}\right)\right]\,,
\end{align}
where the scalar potential is
\begin{equation}
V\,=\,m^2X_0^2-4g^2X_1X_2-4gmX_0\left(X_1+X_2\right)\,,
\end{equation}
and $g$ is the gauge coupling constant, $m$ is the mass parameter, and $\varepsilon_{012345}=+1$. In order to have the supersymmetric $AdS_6$ fixed point from the scalar potential, we have to set $2g=3m$. The norm of form fields are defined by
\begin{equation}
|\omega|^2\,=\,\frac{1}{p!}\omega_{\mu_1\ldots\mu_p}\omega^{\mu_1\ldots\mu_p}\,.
\end{equation}
The equations of motion are presented in appendix \ref{appA2}.

\subsection{The $AdS_2\times\text{disk}\times\text{disk}$ solutions}

Motivated by the $AdS_3\times\text{disk}\times\text{disk}$ solutions in the previous section, we construct $AdS_2\times\text{disk}\times\text{disk}$ solutions of six-dimensional $U(1)^2$-gauged supergravity by trial and error. We also take hints from the $AdS_2\times\text{disk}$ solutions of four-dimensional gauged supergravity and the $AdS_4\times\text{disk}$ solutions of six-dimensional $U(1)^2$-gauged supergravity which we review in appendix \ref{appC} and \ref{appD}, respectively. We only present the obtained solution.

The local form of the $AdS_2\times\text{disk}\times\text{disk}$ solution is
\begin{align} \label{ads2one}
ds_6^2\,=&\,\frac{y^2}{X^2X_1^{2/5}}\left[f^{1/2}\left(\frac{1}{4}ds_{AdS_2}^2+\frac{1}{p}dx^2+\frac{p}{4f}d\psi^2\right)+\frac{y^2X^2}{F}dy^2+\frac{m^2FX^6}{F+y^4X^4}dz^2\right]\,, \notag \\
X_1\,=&\,\left[\frac{m^2y^6}{F+y^4X^4}\right]^{5/8}\,, \qquad X_2\,=\,\left[\frac{m^2y^6}{F+y^4X^4}\right]^{-3/8}\,, \notag \\
A_1\,=&\,-\frac{m^2y^6X^4}{F+y^4X^4}dz\,, \qquad A_2\,=\,\frac{1}{m}\frac{x}{x+s_1}d\psi\,, \notag \\
B\,=&\,-\frac{s_1}{4m}y\,\text{vol}_{AdS_2}\,,
\end{align}
where $q_1$ and $s_1$ are constant parameters. We employed $2g=3m$ and we have
\begin{equation}
F(y)\,=\,m^2y^3\left(y^3+q_1\right)-y^4\,,
\end{equation}
and
\begin{equation}
f(x)\,=\,\left(x+s_1\right)^3x\,, \qquad p(x)\,=\,f(x)-4x^2\,, \qquad X(x)\,=\,\frac{f(x)^{1/4}}{x+s_1}\,.
\end{equation}
Thus the solution has a non-trivial scalar field and two $U(1)$ gauge fields, effectively.

The solution can also be given in the form of
\begin{align}
ds_6^2\,=&\,\frac{y^{5/4}\left(h+yX^4\right)^{1/4}}{m^{1/2}X^2}\left[f^{1/2}\left(\frac{1}{4}ds_{AdS_2}^2+\frac{1}{p}dx^2+\frac{p}{4f}d\psi^2\right)+\frac{X^2}{yh}dy^2+\frac{m^2hX^6}{h+yX^4}dz^2\right]\,, \notag \\
X_1\,=&\,\left[\frac{m^2y^3}{h+yX^4}\right]^{5/8}\,, \qquad X_2\,=\,\left[\frac{m^2y^3}{h+yX^4}\right]^{-3/8}\,, \notag \\
A_1\,=&\,-\frac{m^2y^3X^4}{h+yX^4}dz\,, \qquad A_2\,=\,\frac{1}{m}\frac{x}{x+s_1}d\psi\,, \notag \\
B\,=&\,-\frac{s_1}{4m}y\,\text{vol}_{AdS_2}\,,
\end{align}
where we define
\begin{equation}
h(y)\,=\,m^2\left(y^3+q_1\right)-y\,.
\end{equation}

Now we consider the global completion of the solution, $i.e.$, find the range of the solution where the metric functions are positive definite and the fields are real. We find such solutions when
\begin{equation}
0<y<y_1\,,
\end{equation}
where we define
\begin{align} \label{yone}
y_1\,=&\,\frac{1}{\sqrt{3}m}\left(\tilde{X}+\frac{1}{\tilde{X}}\right)\,, \notag \\
\tilde{X}\,\equiv&\,\frac{1}{\sqrt{3}}\left(\frac{3}{2}\right)^{1/3}\left(-x+\sqrt{x^2-12}\right)^{1/3}\,, \notag \\
x\,\equiv&\,\,9m^3q_1\,,
\end{align}
and $y_1$ is a solution of $h(y)=0$.

Near $y\rightarrow0$, the warp factor vanishes and it is a curvature singularity of the metric,
\begin{equation}
ds_6^2\,\approx\,\frac{q_1^{1/4}y^{5/4}}{X^2}\left[f^{1/2}\left(\frac{1}{4}ds_{AdS_2}^2+\frac{1}{p}dx^2+\frac{p}{4f}d\psi^2\right)+\frac{X^2}{m^2q_1y}dy^2+m^2X^6dz^2\right]\,.
\end{equation}

Approaching $y\rightarrow{y}_1$, the metric becomes to be
\begin{equation}
ds_6^2\,\approx\,\frac{y_1}{m^{1/2}X}\left[f^{1/2}\left(\frac{1}{4}ds_{AdS_2}^2+\frac{1}{p}dx^2+\frac{p}{4f}d\psi^2\right)+\frac{4X^2\left[d\rho^2+\mathcal{E}^2(q_1)\rho^2dz^2\right]}{y_1\left(-h'(y_1)\right)}\right]\,,
\end{equation}
where we introduced a new coordinate, $\rho^2\,=\,y_1-y$. The function, $\mathcal{E}(q_1)$, is given by
\begin{equation} \label{eq1}
\mathcal{E}(q_1)^2\,=\,\frac{m^2y_1h'(y_1)^2X(x_1)^4}{4\left(h+y_1X(x_1)^4\right)}\,,
\end{equation}
where we evaluated it at $x=x_1$ and $y=y_1$ where both disks are at their orbifold singularities. Then, the $y-z$ surface is locally an $\mathbb{R}^2/\mathbb{Z}_\ell$ orbifold if we set
\begin{equation} \label{mathcalE}
\mathcal{E}(q_1)\,=\,\frac{1}{\mathcal{C}\ell}\,=\,\frac{2\pi}{\Delta{z}\ell}\,,
\end{equation}
where $\Delta{z}$ is the period of the coordinate, $z$, and $\ell=1,2,3,\ldots$. Thus the metric on the surface, $\Sigma(y,z)$, has a topology of disk with the origin at $y=y_1$ and the boundary at $y=0$.

We calculate the Euler characteristic of the surface, $\Sigma(y,z)$,
\begin{align}
\chi\left(\Sigma\right)\,=&\,\frac{1}{4\pi}\int_\Sigma{R}_\Sigma\text{vol}_\Sigma\,=\,\frac{2\pi}{4\pi}\frac{y_1^{1/2}\left(2y_1^3-q_1\right)X(x_1)^6}{\left(m^2\left(y_1^3+q_1\right)-y_1+y_1X(x_1)^4\right)^{3/2}}\frac{\Delta{z}}{2\pi} \notag \\
=&\,\mathcal{E}\left(q_1\right)\frac{\Delta{z}}{2\pi}\,=\,\mathcal{C}\mathcal{E}\left(q_1\right)\,=\,\frac{1}{\ell}\,,
\end{align}
which is a natural result for a disk.

\subsection{Uplift to massive type IIA supergravity}

Employing the uplift formula in \cite{Couzens:2022lvg, Faedo:2022rqx}, with the choice of
\begin{equation}
\mu_0\,=\,\cos\xi\cos\theta\,, \qquad \mu_1\,=\,\sin\xi\,, \qquad \mu_2\,=\,\cos\xi\sin\theta\,,
\end{equation}
and $\xi,\theta\in\left[0,\pi/2\right]$, we uplift the $AdS_2\times\text{disk}\times\text{disk}$ solutions to massive type IIA supergravity, \cite{Romans:1985tz}. The metric in the string frame, the dilaton, the Romans mass, and the four-form flux are, respectively,
\begin{align}
ds_{10}^2\,=&\,\frac{\lambda^2\cos^{-1/3}\xi\cos^{-1/3}\theta\tilde{\Delta}^{1/2}y^{1/2}}{X^2} \notag \\
&\times\left[f^{1/2}\left(\frac{1}{4}ds_{AdS_2}^2+\frac{1}{p}dx^2+\frac{p}{4f}d\psi^2\right)+\frac{X^2}{yh}dy^2+\frac{m^2hX^6}{h+yX^4}dz^2\right. \notag \\
& \qquad +\left.\frac{X^2}{g^2y^2}d\xi^2+\frac{X^2}{g^2\tilde{\Delta}y^2}\Big(\left(h+yX^4\right)\sin^2\xi{D}\phi_1^2+y^3\cos^2\xi\left(d\theta^2+\sin^2\theta{D}\phi_2^2\right)\Big)\right]\,, \\
e^\Phi\,=&\,\lambda^2\frac{\left[m^2y^3+\left(m^2q_1-\left(1-X^4\right)y\right)\cos^2\xi\right]^{1/4}}{my^{3/4}}\cos^{-5/6}\xi\cos^{-5/6}\theta\,, \\
F_{(0)}\,=&\,\frac{m}{\lambda^3}\,,
\end{align}
\begin{align} \label{4ads2}
\lambda^{-1}F_{(4)}\,=&\,-\frac{2}{3g^3}\frac{\left[m^2\left(y^3+q_1\right)-\left(1-X^4\right)y\right]\left[5m^2y^3+2\left(m^2q_1-\left(1-X^4\right)y\right)\cos^2\xi\right]}{\left[m^2y^3+\left(m^2q_1-\left(1-X^4\right)y\right)\cos^2\xi\right]^2} \notag \\
& \qquad \times\sin\xi\cos^{7/3}\xi\sin\theta\cos^{1/3}\theta\,\,d\xi\wedge{d}\theta\wedge{D}\phi_1\wedge{D}\phi_2 \notag \\
&-\frac{m^2y^2\left[3m^2q_1-2\left(1-X^4\right)y\right]}{g^3\left[m^2y^3+\left(m^2q_1-\left(1-X^4\right)y\right)\cos^2\xi\right]^2} \notag \\
& \qquad \times\sin^2\xi\cos^{10/3}\xi\sin\theta\cos^{1/3}\theta\,\,dy\wedge{d}\theta\wedge{D}\phi_1\wedge{D}\phi_2 \notag \\
&-\frac{3q_1y^2\left[m^2\left(y^3+q_1\right)-\left(1-X^4\right)y\right]}{g^2\left(y^3+q_1\right)^2\left[m^2y^3+\left(m^2q_1-\left(1-X^4\right)y\right)\cos^2\xi\right]} \notag \\
& \qquad \times\cos^{10/3}\xi\sin\theta\cos^{1/3}\theta{d}y\wedge{d}z\wedge{d}\theta\wedge{d}\phi_2 \notag \\
&-\frac{3m^2s_1y^3}{2g^3\left(x+s_1\right)^2\left[m^2y^3+\left(m^2q_1-\left(1-X^4\right)y\right)\cos^2\xi\right]} \notag \\
& \qquad \times\sin^2\xi\cos^{4/3}\xi\sin\theta\cos^{1/3}\theta{d}x\wedge{d}\psi\wedge{d}\theta\wedge{d}\phi_1 \notag \\ 
&+\frac{3s_1}{2g^3\left(x+s_1\right)^2}\sin\xi\cos^{1/3}\xi\cos^{4/3}\theta{d}x\wedge{d}\psi\wedge{d}\xi\wedge{d}\phi_2 \notag \\ \notag \\
&-\frac{3m}{16g}\frac{s_1yX^2}{f^{1/2}}\cos^{4/3}\xi\cos^{4/3}\theta\,\,dx\wedge{d}\psi\wedge{d}y\wedge{d}z \notag \\
&-\frac{m}{24g}\frac{s_1y^2X^2\left[m^2\left(y^3+q_1\right)-y\right]}{f^{1/2}\left[m^2\left(y^3+q_1\right)-\left(1-X^4\right)y\right]} \notag \\
& \qquad \times\cos^{1/3}\xi\cos^{1/3}\theta\left(\sin\xi\cos\theta{d}\xi+\cos\xi\sin\theta{d}\theta\right)\wedge{d}x\wedge{d}\psi\wedge{d}z\,,
\end{align}
where $D\phi_i\equiv{d}\phi_i-A_i$ and $\phi_1,\phi_2\in\left[0,2\pi\right]$. The parameter, $\lambda$, is introduced in the uplift and will be useful for the flux quantizations. The other fields are trivial. If we set $X=1$ and $s_1=0$, the four-form flux reduces back to that of the $AdS_4\times\text{disk}$ solutions, \eqref{4ads4}.

\subsection{Uplifted metric}

\begin{figure}[t] 
\begin{center} 
\includegraphics[width=4.5in]{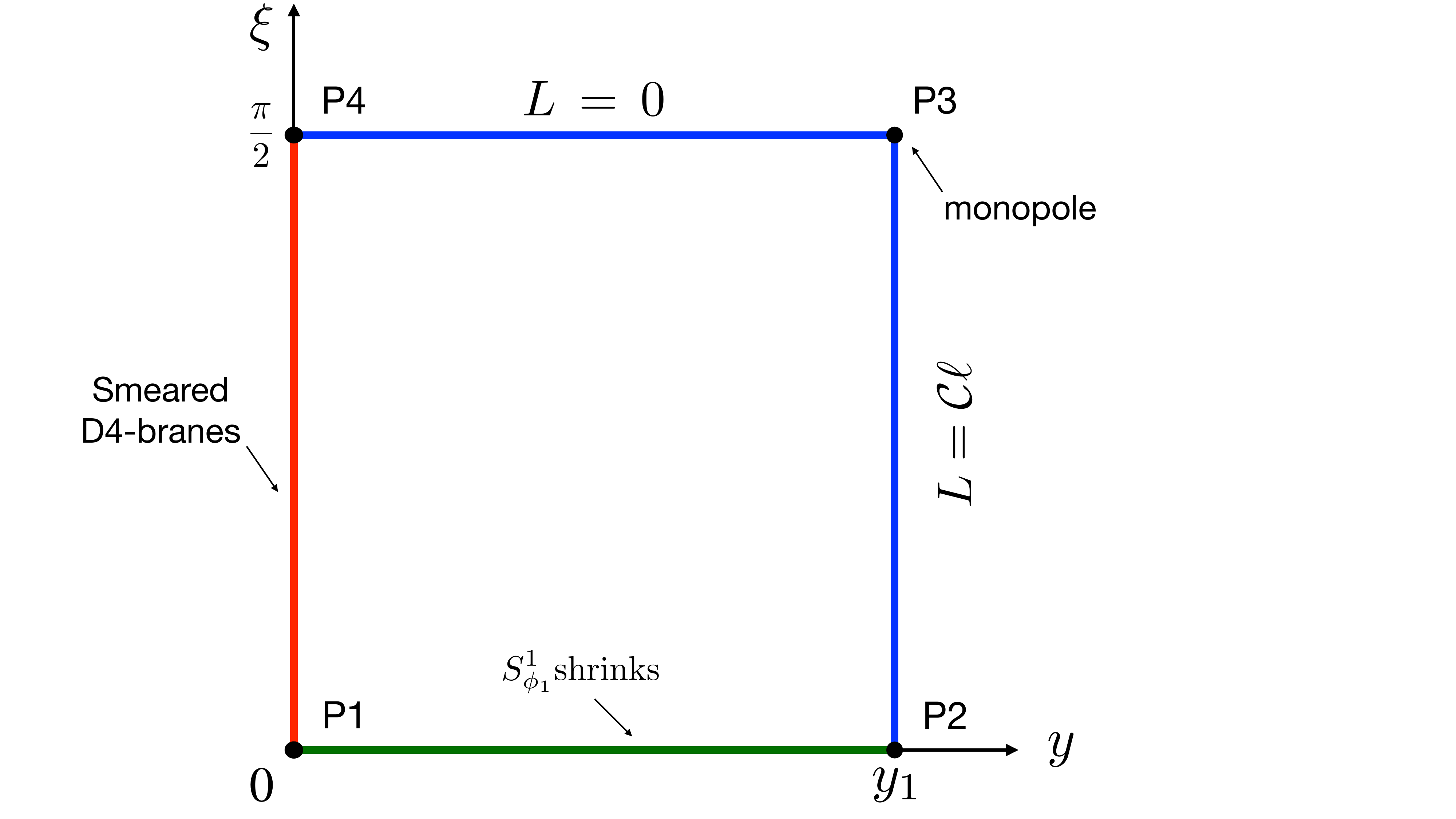} 
\caption{{\it The two-dimensional base space, $B_2$, spanned by $y$ and $\xi$.}} \label{fig2}
\end{center}
\end{figure}

The six-dimensional internal space of the uplifted metric is an $S_z^1\,\times\,S^3$ fibration over the 2d base space, $B_2$, of $(y,\xi)$. The 2d base space is a rectangle of $(y,\xi)$ over $[0,y_1)\,\times\left[0,\frac{\pi}{2}\right]$. See Figure \ref{fig2}. We explain the geometry of the internal space by three regions of the 2d base space, $B_2$.

\begin{itemize}
\item Region I: The side of $\mathsf{P}_1\mathsf{P}_2$.
\item Region II: The sides of $\mathsf{P}_2\mathsf{P}_3$ and $\mathsf{P}_3\mathsf{P}_4$.
\item Region III: The side of $\mathsf{P}_1\mathsf{P}_4$.
\end{itemize}

\noindent {\bf Region I:} On the side of $\xi\,=\,0$, the $\phi_1$ circle, $S^1_{\phi_1}$, shrinks.

\bigskip

\noindent {\bf Region II: Monopole} In order to observe the property of the 2d base, \eqref{2db}, we fix the gauge of the six-dimensional gauge field to be
\begin{equation}
A_1\,=\,-\frac{m^2y^6X^4}{F+y^4X^4}dz-\frac{1}{3}\,.
\end{equation}
We break $D\phi_1^2=\left(d\phi_1-A_1\right)^2$ and complete the square of $dz$ to obtain the metric of
\begin{align}
ds_{10}^2\,=&\,\frac{\lambda^2\cos^{-1/3}\xi\cos^{-1/3}\theta\tilde{\Delta}^{1/2}y^{1/2}}{X^2} \notag \\
&\times\left[f^{1/2}\left(\frac{1}{4}ds_{AdS_2}^2+\frac{1}{p}dx^2+\frac{p}{4f}d\psi^2\right)+\frac{X^2}{yh}dy^2\right. \notag \\
& \qquad +\frac{X^2}{g^2y^2}d\xi^2+\frac{X^2}{g^2\tilde{\Delta}y^2}y^3\cos^2\xi\left(d\theta^2+\sin^2\theta{D}\phi_2^2\right)\,, \\
& \qquad +R_z^2\left(dz-gLd\phi_1\right)^2+R_{\phi_1}^2d\phi_1^2\Big]\,,
\end{align}
where we define
\begin{align}
R_z^2\,=&\,\frac{X^2Y}{9g^2y^2\left(h+yX^4\right)\tilde{\Delta}}\,, \notag \\
R_{\phi_1}^2\,=&\,\frac{9m^2h\left(h+yX^4\right)X^6\sin^2\xi}{Y}\,, \notag \\
L\,=&\,\frac{3\left(h+yX^4\right)\left(h+y\left(1-3m^2y^2\right)X^4\right)\sin^2\xi}{gY}\,,
\end{align}
with
\begin{align}
Y\,\equiv\,&9g^2m^2hy^2\left(h+yX^4\right)X^4\cos^2\xi \notag \\
&+\left(h^2+y^2\left(1-3m^2y^2\right)^2X^8+hy\left(2-3m^2\left(2y^2-3g^2y^4\right)\right)X^4\right)\sin^2\xi\,.
\end{align}

The function, $L(y,\xi)$, is piecewise constant along the sides of $y=y_1$ and $\xi=\frac{\pi}{2}$ of the 2d base, $B_2$, 
\begin{equation} \label{2db}
L\left(y,\frac{\pi}{2}\right)\,=\,0\,, \qquad L(y_1,\xi)\,=\,\frac{1}{\mathcal{E}(q_1)}\,=\,\mathcal{C}\ell\,=\,\frac{\Delta{z}}{2\pi}\ell\,.
\end{equation}
The jump at the corner, $(y,\xi)=\left(y_1,\frac{\pi}{2}\right)$, indicates a monopole source for the $D\phi_1$ fibration.

\bigskip

\noindent {\bf Region III: Smeared D4-D8-branes} We introduce a new parametrization of the coordinate, $r\equiv{y}^{3/2}$. In the limit, $r\rightarrow0$, the metric asymptotes to
\begin{align}
ds_{10}^2\,\approx&\,\frac{\lambda^2q_1^{1/2}\cos^{2/3}\xi\cos^{-1/3}\theta{r}^{-1/3}}{X^2}\left[r^{2/3}\left(f^{1/2}\left(\frac{1}{4}ds_{AdS_2}^2+\frac{1}{p}dx^2+\frac{p}{4f}d\psi^2\right)+m^2X^6dz^2\right)\right. \notag \\
&+\left.\frac{X^2}{g^2q_1}r^{-2/3}\Big(q_1\left(d\xi^2+\tan^2\xi{D}\phi_1^2\right)+dr^2+r^2\left(d\theta^2+\sin^2\theta{D}\phi_2^2\right)\Big)\right]\,.
\end{align}
The metric implies the smeared D4-D8-brane sources. The D4-D8-branes are 
\begin{itemize}
\item extended along the $AdS_2$, $x$, $\psi$, and $z$ directions;
\item localized at the origin of the $\mathbb{R}^3$ parametrized by $S^2$ and $r$, $ds^2_{\mathbb{R}^3}=dr^2+r^2ds^2_{S^2}$;
\item smeared along the $\xi$ and $\phi_1$ directions. 
\end{itemize}
This matches the metric of the smeared D4-D8-branes obtained in the appendix B of \cite{Bah:2017wxp}.

\subsection{Flux quantization}

For the flux quantization, we follow the analysis in \cite{Suh:2021aik}. 

The quantization condition on the Romans mass is given by
\begin{equation}
\left(2\pi\ell_s\right)F_{(0)}\,\equiv\,n_0\,\in\,\mathbb{Z}\,,
\end{equation}
where $\ell_s$ is the string length, $n_0=8-N_f$ and $N_f$ is the number of D8-branes. 

The integral of four-form flux through any four-cycle is an integer,
\begin{equation} \label{romansq}
\frac{1}{\left(2\pi\ell_s\right)^3}\int_{M_4}F_{(4)}\,\in\,\mathbb{Z}\,.
\end{equation}
First, from the flux through the four-cylce of $\xi\theta\phi_1\phi_2$, we obtain
\begin{equation} \label{four1q}
\frac{1}{\left(2\pi\ell_s\right)^3}\int_{\xi\theta\phi_1\phi_2}{F}_{(4)}\,=\,\frac{3\lambda}{8\pi\ell_s^3g^3}\,\equiv\,N\,,
\end{equation}
where $N\in\mathbb{N}$ is the number of D4-branes wrapping the disk$\times$disk. From \eqref{romansq} and \eqref{four1q}, we find, \cite{Faedo:2021nub},
\begin{equation} \label{fluxglam}
g^8\,=\,\frac{1}{\left(2\pi\ell_s\right)^8}\frac{18\pi^6}{N^3n_0}\,, \qquad \lambda^8\,=\,\frac{8\pi^2}{9Nn_0^3}\,.
\end{equation}

Second, from the flux through the four cycle of $yz\theta\phi_2$ including the disk, $\Sigma(y,z)$, at $\xi=0$, we find
\begin{equation} \label{qK}
\frac{1}{\left(2\pi\ell_s\right)^3}\int_{yz\theta\phi_2}{F}_{(4)}\,=\,-\frac{1}{\left(2\pi\ell_s\right)^3}\frac{3\lambda}{4g^2}\frac{y_1^3}{y_1^3+q_1}2\pi\Delta{z}\,=\,-\frac{N}{\ell}\frac{1}{\mathcal{E}(q_1)}\frac{gy_1^3}{y_1^3+q_1}\,\equiv\,-K\,,
\end{equation}
where $K\in\mathbb{Z}$ is another integer. By eliminating $\mathcal{E}(q_1)$ from the constraints, \eqref{mathcalE} and \eqref{qK}, and with the expression of $y_1(q_1)$ in \eqref{yone}, we obtain
\begin{equation} \label{q1one}
q_1\,=\,\frac{K^{1/2}\left(g\,\mathcal{C}N-K\right)}{m^3\left(g\,\mathcal{C}N\right)^{3/2}}\,,
\end{equation}
and we also find
\begin{equation} \label{yoneyone}
y_1\,=\,\frac{1}{m}\sqrt{\frac{K}{g\,\mathcal{C}N}}\,.
\end{equation}
Then, by plugging $y_1(q_1)$, \eqref{yoneyone}, in \eqref{mathcalE} with the expression of $\mathcal{E}(q_1)$ in \eqref{eq1}, we also find another expression for $q_1$,
\begin{equation} \label{q1two}
q_1\,=\,\frac{K^{1/2}\left(9m^2\mathcal{C}K^2\ell^2-4gNK-6gm^2\mathcal{C}^2NK\ell^2+gm^2\mathcal{C}^3N^2\ell^2\right)}{4g^{5/2}m^3\mathcal{C}^{3/2}N^{5/2}}\,.
\end{equation}
Finally, identifying \eqref{q1one} and \eqref{q1two}, we can solve for $\mathcal{C}$ and then for $q_1$ in terms of the quantum numbers, $N$, $K$, and $\ell$,
\begin{equation}
\mathcal{C}\,=\,\frac{3\left(N+K\ell\right)}{gN\ell}\,, \qquad q_1\,=\,\frac{1}{3\sqrt{3}m^3}\frac{\sqrt{K\ell}\left(2K\ell+3N\right)}{\left(K\ell+N\right)^{3/2}}\,.
\end{equation}
It is amazing that the dependence on $X$ disappears all along.

Third, from the flux through the four cycle of $x\psi\xi\phi_1$ including the disk, $\Sigma(x,\psi)$, at $\theta=0$, we find
\begin{equation} \label{qK3}
\frac{1}{\left(2\pi\ell_s\right)^3}\int_{x\psi\xi\phi_1}{F}_{(4)}\,=\,-\frac{1}{\left(2\pi\ell_s\right)^3}\frac{9\lambda}{8g^3}\frac{x_1}{x_1+s_1}2\pi\Delta\psi\,=\,\frac{N}{k}\frac{1}{\mathcal{E}(s_1)}\frac{x_1}{x_1+s_1}\,\equiv\,M\,,
\end{equation}
where $M\in\mathbb{Z}$ is another integer and the orbifold number, $k$, and the function, $\mathcal{E}(s_1)$, are introduced in appendix \ref{appC}. By eliminating $\mathcal{E}(s_1)$ from the constraints, \eqref{mathcalE3} and \eqref{qK3}, and with the expression of $x_1(s_1)$ in \eqref{yone3}, we obtain
\begin{equation} \label{s1one}
s_1\,=\,\frac{2K^{1/2}\left(g\mathcal{D}N-M\right)}{\left(g\mathcal{D}M\right)^{3/2}}\,,
\end{equation}
and we also find
\begin{equation} \label{xonexone}
x_1\,=\,2\left(\frac{M}{g\mathcal{D}N}\right)^{3/2}\,.
\end{equation}
The flux quantiztions on the four cycles, $yz\theta\phi_2$ in \eqref{qK} and $x\psi\xi\phi_1$ in \eqref{qK3}, effectively reduce to flux quantizations of six-dimensional field strengths, $F_1$ and $F_2$, respectively. Then, by plugging $s_1$, \eqref{s1one}, and $x_1$, \eqref{xonexone}, in $\mathcal{E}(s_1)$, \eqref{eq13}, we also find $\mathcal{D}$ and then $s_1$,
\begin{equation}
\mathcal{D}\,=\,\frac{2gN+3Mk}{gNk}\,, \qquad s_1\,=\,\frac{4\sqrt{Mk}\left(gN+Mk\right)}{\left(2gN+3Mk\right)^{3/2}}\,.
\end{equation}

\subsection{The Bekenstein-Hawking entropy}

For the metric in the string frame,
\begin{equation}
ds_{10}^2\,=\,e^{2\mathcal{A}}\left(ds_{AdS_2}^2+ds_{M_8}^2\right)\,,
\end{equation}
the Bekenstein-Hawking entropy of the presumed black hole is given by, $e.g.$, \cite{Faedo:2021nub},
\begin{equation}
S_{\text{BH}}\,=\frac{1}{4G_N^{(2)}}\,=\,\frac{8\pi^2}{\left(2\pi\ell_s\right)^8}\int_{M_8}e^{8\mathcal{A}-2\Phi}\text{vol}_{M_8}\,.
\end{equation}
From above, we obtain the Bekenstein-Hawking entropy,
\begin{equation} \label{holcen2}
S_{\text{BH}}\,=\,\frac{8\pi^2}{\left(2\pi\ell_s\right)^8}\frac{3\pi^2\lambda^4}{20g^4}x_1\Delta\psi\,y_1^3\Delta{z}\,=\,\frac{2\sqrt{6}\pi}{5\sqrt{8-N_f}}\sqrt{\frac{N^3K^3\ell}{N+K\ell}}\sqrt{\frac{M^3k}{g^2N^2\left(2gN+3Mk\right)}}\,,
\end{equation}
where the second and third factors are from the disks, $\Sigma(y,z)$ and $\Sigma(x,\psi)$, respectively. The Bekenstein-Hawking entropy can also be written by
\begin{equation}
S_{\text{BH}}\,=\,\sqrt{\frac{M^3k}{g^2N^2\left(2gN+3Mk\right)}}\mathcal{F}\,,
\end{equation}
where $\mathcal{F}$ is the holographic free energy of $AdS_4\times\text{disk}$ solutions in \eqref{hfree}. Note that if $N\sim{K}\sim{M}$, it scales as $N^{5/2}$, as the theories originating from the compactifications of 5d Seiberg theories.

\section{D4-branes wrapped on spindle $\ltimes$ disk} \label{sec5}

The local form of the $AdS_2\times\text{spindle}\ltimes\text{disk}$ solution can be easily obtained from \eqref{ads2one} by $dz\rightarrow{D}z$ and recovering $s_2$ as it is in \eqref{m2one}. As the changes do not affect most of the analysis on the $AdS_2\times\text{disk}\times\text{disk}$ solution in the previous section, we will be brief. The local form of the $AdS_2\times\text{spindle}\ltimes\text{disk}$ solution is
\begin{align} \label{ads2one}
ds_6^2\,=&\,\frac{y^2}{X^2X_1^{2/5}}\left[f^{1/2}\left(\frac{1}{4}ds_{AdS_2}^2+\frac{1}{p}dx^2+\frac{p}{4f}d\psi^2\right)+\frac{y^2X^2}{F}dy^2+\frac{m^2FX^6}{F+y^4X^4}Dz^2\right]\,, \notag \\
X_1\,=&\,\left[\frac{m^2y^6}{F+y^4X^4}\right]^{5/8}\,, \qquad X_2\,=\,\left[\frac{m^2y^6}{F+y^4X^4}\right]^{-3/8}\,, \notag \\
A_1\,=&\,-\frac{m^2y^6X^4}{F+y^4X^4}Dz\,, \qquad A_2\,=\,\frac{1}{m}\frac{x}{x+s_1}d\psi\,, \notag \\
B\,=&\,-\frac{s_1}{4m}y\,\text{vol}_{AdS_2}\,,
\end{align}
where $q_1$, $s_1$, and $s_2$ are constant parameters. We employed $2g=3m$ and we have
\begin{equation}
F(y)\,=\,m^2y^3\left(y^3+q_1\right)-y^4\,,
\end{equation}
and
\begin{equation}
f(x)\,=\,\left(x+s_1\right)^3\left(x+s_2\right)\,, \qquad p(x)\,=\,f(x)-4x^2\,, \qquad X(x)\,=\,\frac{f(x)^{1/4}}{x+s_1}\,,
\end{equation}
with
\begin{equation}
Dz\,=\,dz+A_2\,.
\end{equation}
Thus the solution has a non-trivial scalar field and two $U(1)$ gauge fields, effectively. The spindle can be either in the twist class or in the anti-twist class depending on the choice of the parameters, \cite{Ferrero:2021etw}.

The solution can also be given in the form of
\begin{align}
ds_6^2\,=&\,\frac{y^{5/4}\left(h+yX^4\right)^{1/4}}{m^{1/2}X^2}\left[f^{1/2}\left(\frac{1}{4}ds_{AdS_2}^2+\frac{1}{p}dx^2+\frac{p}{4f}d\psi^2\right)+\frac{X^2}{yh}dy^2+\frac{m^2hX^6}{h+yX^4}Dz^2\right]\,, \notag \\
X_1\,=&\,\left[\frac{m^2y^3}{h+yX^4}\right]^{5/8}\,, \qquad X_2\,=\,\left[\frac{m^2y^3}{h+yX^4}\right]^{-3/8}\,, \notag \\
A_1\,=&\,-\frac{m^2y^3X^4}{h+yX^4}Dz\,, \qquad A_2\,=\,\frac{1}{m}\frac{x}{x+s_1}d\psi\,, \notag \\
B\,=&\,-\frac{s_1}{4m}y\,\text{vol}_{AdS_2}\,,
\end{align}
where we define
\begin{equation}
h(y)\,=\,m^2\left(y^3+q_1\right)-y\,.
\end{equation}
The solution reduces back to the local form of $AdS_2\times\text{disk}\times\text{disk}$ solution in the previous section by
\begin{equation}
Dz\,\rightarrow{d}z\,, \qquad s_1\,\rightarrow\,0\,.
\end{equation}

Similar to \eqref{holcen2}, we find the Bekenstein-Hawking entropy to be
\begin{align}
S_{\text{BH}}\,=\,\frac{8\pi^2}{\left(2\pi\ell_s\right)^8}\frac{3\pi^2\lambda^4}{20g^4}\left(x_2-x_1\right)\Delta\psi\,y_1^3\Delta{z}\,.
\end{align}
However, the expression of Bekenstein-Hawking entropy for the $AdS_2\times\text{spindle}$ solution is quite unwieldy, \cite{Couzens:2021cpk}, and we refrain from further calculation of the Bekenstein-Hawking entropy of the solution.

\section{Conclusions} \label{sec6}

In this paper, we constructed and studied the $AdS_3\times\text{disk}\times\text{disk}$ and $AdS_2\times\text{disk}\times\text{disk}$ solutions of $U(1)^2$-gauged supergravity in seven and six dimensions, respectively. We uplifted the solutions to eleven-dimensional and massive type IIA supergravity, respectively, and studied the disk geometry of the solutions. We performed the flux quantizations and calculated the holographic central charge and the Bekenstein-Hawking entropy, respectively. In a similar manner, we presented the $AdS_3\times\text{spindle}\ltimes\text{disk}$ and $AdS_2\times\text{spindle}\ltimes\text{disk}$ solutions.

As it was discussed in the introduction, we have constructed the $AdS_{2,3}\times\text{disk}\times\text{disk}$ solutions embedded in the maximal $AdS_{4,5}\times\text{disk}$ solutions, respectively. It is an open question if there are $AdS_{2,3}\times\text{disk}\times\text{disk}$ solutions embedded in the minimal $AdS_{4,5}\times\text{disk}$ solutions.

The $AdS_2\times\text{disk}\times\text{disk}$ solution is analogous to the $AdS_2\times\Sigma_\mathfrak{g}\times\Sigma_\mathfrak{g}$ solution where $\Sigma_\mathfrak{g}$ is a Riemann surface of genus, $\mathfrak{g}$, \cite{Suh:2018tul, Hosseini:2018usu, Suh:2018szn, Kim:2019fsg}. The Bekenstein-Hawking entropy of those solutions were precisely reproduced by the topologically twisted index, \cite{Hosseini:2018uzp, Crichigno:2018adf}, of dual 5d gauge theories, \cite{Brandhuber:1999np, Seiberg:1996bd}. It would be most interesting if we could microscopically count the Bekenstein-Hawking entropy of the black hole with $AdS_2\times\text{disk}\times\text{disk}$ horizon.

\bigskip
\bigskip
\leftline{\bf Acknowledgements}
\noindent We are grateful to Christopher Couzens for explaining his work, \cite{Bomans:2023ouw}, for us. This work was supported by the Kumoh National Institute of Technology.

\appendix
\section{The equations of motion} \label{appA}
\renewcommand{\theequation}{A.\arabic{equation}}
\setcounter{equation}{0} 

\subsection{$U(1)^2$-gauged supergravity in seven dimensions} \label{appA1}

We present the equations of motion  derived from the Lagrangian in \eqref{sevenlag},
\begin{align}
R_{\mu\nu}\,=\,&6\partial_\mu\lambda_1\partial_\nu\lambda_1+6\partial_\mu\lambda_2\partial_\nu\lambda_2+8\partial_{(\mu}\lambda_1\partial_{\nu)}\lambda_2+\frac{1}{5}g_{\mu\nu}V \notag \\
&+\frac{1}{2}e^{-4\lambda_1}\left(F_{\mu\rho}^{12}F_\nu^{12\rho}-\frac{1}{10}g_{\mu\nu}F_{\rho\sigma}^{12}F^{12\rho\sigma}\right)+\frac{1}{2}e^{-4\lambda_2}\left(F_{\mu\rho}^{34}F_\nu^{34\rho}-\frac{1}{10}g_{\mu\nu}F_{\rho\sigma}^{34}F^{34\rho\sigma}\right) \notag \\
&+\frac{1}{4}e^{-4\lambda_1-4\lambda_2}\left(S_{\mu\rho\sigma}^5S_\nu^{5\rho\sigma}-\frac{2}{15}g_{\mu\nu}S_{\rho\sigma\delta}^5S^{5\rho\sigma\delta}\right)\,,
\end{align}
\begin{align}
\frac{1}{\sqrt{-g}}\partial_\mu\left(\sqrt{-g}g^{\mu\nu}\partial_\nu\left(3\lambda_1+2\lambda_2\right)\right)+\frac{1}{4}e^{-4\lambda_1}F_{\mu\nu}^{12}F^{12\mu\nu}+\frac{1}{12}e^{-4\lambda_1-4\lambda_2}S_{\mu\nu\rho}^5S^{5\mu\nu\rho}-\frac{g^2}{4}\frac{\partial{V}}{\partial\lambda_1}\,=\,0\,, \notag \\
\frac{1}{\sqrt{-g}}\partial_\mu\left(\sqrt{-g}g^{\mu\nu}\partial_\nu\left(2\lambda_1+3\lambda_2\right)\right)+\frac{1}{4}e^{-4\lambda_2}F_{\mu\nu}^{34}F^{34\mu\nu}+\frac{1}{12}e^{-4\lambda_1-4\lambda_2}S_{\mu\nu\rho}^5S^{5\mu\nu\rho}-\frac{g^2}{4}\frac{\partial{V}}{\partial\lambda_2}\,=\,0\,, 
\end{align}
\begin{align}
d\left(e^{-4\lambda_1}*F^{12}\right)+e^{-4\lambda_1-4\lambda_2}*S^5\wedge\,F^{34}\,=\,0\,, \notag \\
d\left(e^{-4\lambda_2}*F^{34}\right)+e^{-4\lambda_1-4\lambda_2}*S^5\wedge\,F^{12}\,=\,0\,, \notag \\
dS^5-ge^{-4\lambda_1-4\lambda_2}*S^5-F^{12}\wedge\,F^{34}\,=\,0\,.
\end{align}

\subsection{$U(1)^2$-gauged supergravity in six dimensions} \label{appA2}

We present the equations of motion  derived from the action in \eqref{actionaction},
\begin{align}
&R_{\mu\nu}-\frac{1}{2}\sum_{i=1}^2\partial_\mu\varphi_i\partial_\nu\varphi_i-\frac{1}{4}Vg_{\mu\nu}-\frac{1}{2}\sum_{i=1}^2X_i^{-2}\left(F_{i\mu\rho}F_{i\nu}\,^\rho-\frac{1}{8}g_{\mu\nu}F_{i\rho\sigma}F_i\,^{\rho\sigma}\right) \notag \\
-&\frac{m^2}{4}\left(X_1X_2\right)^{-1}\left(B_{\mu\rho}B_\nu\,^\rho-\frac{1}{8}g_{\mu\nu}B_{\rho\sigma}B^{\rho\sigma}\right)-\frac{1}{16}\left(X_1X_2\right)^2\left(H_{\mu\rho\sigma}H_\nu\,^{\rho\sigma}-\frac{1}{6}g_{\mu\nu}H_{\rho\sigma\lambda}H^{\rho\sigma\lambda}\right)\,=\,0\,,
\end{align}
\begin{align}
\frac{1}{\sqrt{-g}}\partial_\mu\left(\sqrt{-g}g^{\mu\nu}\partial_\nu\varphi_1\right)\,-&\,\frac{\partial{V}}{\partial\varphi_1}-\frac{1}{2\sqrt{2}}X_1^{-2}F_{1\mu\nu}F_1\,^{\mu\nu}+\frac{1}{2\sqrt{2}}X_2^{-2}F_{2\mu\nu}F_2\,^{\mu\nu}\,=\,0\,, \notag \\
\frac{1}{\sqrt{-g}}\partial_\mu\left(\sqrt{-g}g^{\mu\nu}\partial_\nu\varphi_2\right)\,-&\,\frac{\partial{V}}{\partial\varphi_2}-\frac{1}{4\sqrt{2}}X_1^{-2}F_{1\mu\nu}F_1\,^{\mu\nu}-\frac{1}{4\sqrt{2}}X_2^{-2}F_{2\mu\nu}F_2\,^{\mu\nu} \notag \\
-&\frac{m^2}{8\sqrt{2}}\left(X_1X_2\right)^{-1}B_{\mu\nu}B^{\mu\nu}+\frac{1}{24\sqrt{2}}\left(X_1X_2\right)^2H_{\mu\nu\rho}H^{\mu\nu\rho}\,=\,0\,, 
\end{align}
\begin{align}
\mathcal{D}_\nu\left(X_1^{-2}F_1^{\nu\mu}\right)\,=&\,\frac{1}{24}\sqrt{-g}\varepsilon^{\mu\nu\rho\sigma\tau\lambda}F_{2\nu\rho}H_{\sigma\tau\lambda}\,, \notag \\
\mathcal{D}_\nu\left(X_2^{-2}F_2^{\nu\mu}\right)\,=&\,\frac{1}{24}\sqrt{-g}\varepsilon^{\mu\nu\rho\sigma\tau\lambda}F_{1\nu\rho}H_{\sigma\tau\lambda}\,, \notag \\
\mathcal{D}_\nu\left(\left(X_1X_2\right)^{-1}B^{\nu\mu}\right)\,=&\,\frac{1}{24}\sqrt{-g}\varepsilon^{\mu\nu\rho\sigma\tau\lambda}B_{\nu\rho}H_{\sigma\tau\lambda}\,, \notag \\
\mathcal{D}_\rho\left(\left(X_1X_2\right)^2H^{\rho\nu\mu}\right)\,=&\,-\frac{1}{4}\sqrt{-g}\varepsilon^{\mu\nu\rho\sigma\tau\lambda}\left(\frac{m^2}{2}B_{\rho\sigma}B_{\tau\lambda}+F_{i\rho\sigma}F_{i\tau\lambda}\right)-2m^2\left(X_1X_2\right)^{-1}B^{\mu\nu}\,.
\end{align}

\section{D3-branes wrapped on a spindle and on a disk} \label{appB}
\renewcommand{\theequation}{B.\arabic{equation}}
\setcounter{equation}{0} 

The $AdS_3\times\text{disk}$ solutions form D3-branes wrapped on a disk were studied in \cite{Couzens:2021tnv, Suh:2021ifj}. The local form of the solution can also be obtained from the $AdS_3\times\text{spindle}$ solutions, \cite{Boido:2021szx, Ferrero:2021etw}, by taking a special limit. We review the $AdS_3\times\text{disk}$ solutions in the conventions of \cite{Ferrero:2021etw}.

The local form of the $AdS_3\times\text{spindle}$ solution is given by, \cite{Ferrero:2021etw},
\begin{align} \label{d3one}
ds_5^2\,&=\,f^{1/2}\left[ds_{AdS_3}^2+\frac{1}{4p}dx^2+\frac{p}{f}d\psi^2\right]\,, \notag \\
A^I\,&=\,\frac{x}{f_I}d\psi\,, \qquad X^I\,=\,\left(\frac{f_1f_2f_3}{f_I^3}\right)^{1/3}\,,
\end{align}
where we define
\begin{equation}
f_I\,=\,x+s_I\,, \qquad f\,=\,f_1f_2f_3\,, \qquad p\,=\,f-x^2\,,
\end{equation}
and $s_I$, $I=1,2,3$, are constants. It is a solution of $U(1)^3$-gauged supergravity in five dimensions, also known as the STU model. 

The $AdS_3\times\text{spindle}$ solution is globally defined in the range, $x\in[x_1,x_2]$, where $x_1$ and $x_2$ are functions of the orbifold numbers, $n_+$ and $n_-$, at the north and south poles. Then one can find the expression of, \cite{Boido:2021szx},
\begin{equation} \label{x2x1dphi}
\left(x_2-x_1\right)\frac{\Delta\phi}{2\pi}\,=\,\frac{q^2\left(n_--n_+-2q\right)}{n_-n_+\left[n_-\left(n_++2q\right)-q\left(2n_++3q\right)\right]}\,,
\end{equation}
which we employ in the main body of the paper in the calculation of holographic central charge. For the details of the analysis of the solution, we refer the reader to \cite{Boido:2021szx, Ferrero:2021etw}. 

For the $AdS_3\times\text{disk}$ solutions, we consider the special case of the T$^3$ model, $i.e.$, $s_1=s_2$ and $s_3=0$, and the local form of the solution is given by
\begin{align}
ds_5^2\,&=\,f^{1/2}\left[ds_{AdS_3}^2+\frac{1}{4p}dx^2+\frac{p}{f}d\psi^2\right]\,, \notag \\
X^1\,&=\,X^2\,=\,\frac{f^{1/3}}{x+s_1}\,, \qquad X^3\,=\,\frac{f^{1/3}}{x}\,, \notag \\
A^1\,&=A^2\,=\,\frac{x}{x+s_1}d\psi\,, \qquad A^3\,=\,\text{trivial}\,,
\end{align}
and we have
\begin{equation}
f\,=\,\left(x+s_1\right)^2x\,, \qquad p\,=\,f-x^2\,.
\end{equation}

Now we consider the global completion of the solution, $i.e.$, find the range of the solution where the metric functions are positive definite and the fields are real. We find such solutions when
\begin{equation}
0<x<x_1\,,
\end{equation}
where we find
\begin{equation} \label{yone4}
x_1\,=\,\frac{1}{2}+s_1-\frac{1}{2}\sqrt{1+4s_1}\,,
\end{equation}
and $x_1$ is a solution of $p(x)=0$.


Approaching $x\rightarrow{x}_1$, the metric becomes to be
\begin{equation}
ds_5^2\,\approx\,\left(x_1+s_1\right)x_1^{1/2}\left[ds_{AdS_3}^2+\frac{d\sigma^2+\mathcal{E}^2(s_1)\sigma^2d\psi^2}{-p'(x_1)}\right]\,,
\end{equation}
where we introduced a new coordinate, $\sigma^2\,=\,x_1-x$. The function, $\mathcal{E}(s_1)$, is given by
\begin{equation} \label{eq14}
\mathcal{E}(s_1)\,=\,\sqrt{1+4s_1}\,.
\end{equation}
Then, the $x-\psi$ surface is locally an $\mathbb{R}^2/\mathbb{Z}_\ell$ orbifold if we set
\begin{equation} \label{mathcalE4}
\mathcal{E}(s_1)\,=\,\frac{1}{\mathcal{D}k}\,=\,\frac{2\pi}{\Delta\psi{k}}\,,
\end{equation}
where $\Delta\psi$ is the period of the coordinate, $\psi$, and $k=1,2,3,\ldots$. Thus the metric on the surface, $\Sigma(x,\psi)$, has a topology of disk with the origin at $x=x_1$ and the boundary at $x=0$.

We calculate the Euler characteristic of the surface, $\Sigma(x,\psi)$,
\begin{equation}
\chi\left(\Sigma\right)\,=\,\frac{1}{4\pi}\int_\Sigma{R}_\Sigma\text{vol}_\Sigma\,=\,\mathcal{E}\left(s_1\right)\frac{\Delta\psi}{2\pi}\,=\,\mathcal{D}\mathcal{E}\left(s_1\right)\,=\,\frac{1}{k}\,,
\end{equation}
which is a natural result for a disk.

Further analysis of the disk solutions are given in \cite{Couzens:2021tnv, Suh:2021ifj} in different coordinate systems.

\section{M2-branes wrapped on a spindle and on a disk} \label{appC}
\renewcommand{\theequation}{C.\arabic{equation}}
\setcounter{equation}{0} 

The $AdS_2\times\text{disk}$ solutions form M2-branes wrapped on a disk were studied in \cite{Suh:2021hef, Couzens:2021rlk}. The local form of the solution can also be obtained from the $AdS_2\times\text{spindle}$ solutions, \cite{Couzens:2021rlk, Ferrero:2021etw, Couzens:2021cpk}, by taking a special limit. We review the $AdS_2\times\text{disk}$ solutions in the conventions of \cite{Ferrero:2021etw}.

The local form of the $AdS_2\times\text{spindle}$ solution is given by, \cite{Ferrero:2021etw},
\begin{align} \label{m2one}
ds_4^2\,&=\,f^{1/2}\left[\frac{1}{4}ds_{AdS_2}^2+\frac{1}{p}dx^2+\frac{p}{4f}d\psi^2\right]\,, \notag \\
A^I\,&=\,\frac{x}{f_I}d\psi\,, \qquad X^I\,=\,\left(\frac{f_1f_2f_3f_4}{f_I^4}\right)^{1/4}\,,
\end{align}
where we define
\begin{equation}
f_I\,=\,x+s_I\,, \qquad f\,=\,f_1f_2f_3f_4\,, \qquad p\,=\,f-4x^2\,,
\end{equation}
and $s_I$, $I=1,2,3,4$, are constants. It is a solution of $U(1)^4$-gauged supergravity in four dimensions, also known as the STU model.

For the $AdS_2\times\text{disk}$ solutions, we consider the special case of the T$^3$ model, $i.e.$, $s_1=s_2=s_3$ and $s_4=0$, and the local form of the solution is given by
\begin{align}
ds_4^2\,&=\,f^{1/2}\left[\frac{1}{4}ds_{AdS_2}^2+\frac{1}{p}dx^2+\frac{p}{4f}d\psi^2\right]\,, \notag \\
X^1\,&=\,X^2\,=\,X^3\,=\,\frac{f^{1/4}}{x+s_1}\,, \qquad X^4\,=\,\frac{f^{1/4}}{x}\,, \notag \\
A^1\,&=A^2\,=A^3\,=\,\frac{x}{x+s_1}d\psi\,, \qquad A^4\,=\,\text{trivial}\,,
\end{align}
and we have
\begin{equation}
f\,=\,\left(x+s_1\right)^3x\,, \qquad p\,=\,f
-4x^2\,.
\end{equation}

Now we consider the global completion of the solution, $i.e.$, find the range of the solution where the metric functions are positive definite and the fields are real. We find such solutions when
\begin{equation}
0<x<x_1\,,
\end{equation}
where we define
\begin{align} \label{yone3}
x_1\,=&\,\frac{1}{3}\left(-3s_1+12\tilde{X}+\frac{1}{\tilde{X}}\right)\,, \notag \\
\tilde{X}\,\equiv&\,\frac{1}{2\,6^{2/3}}\left(-\tilde{x}+\sqrt{\tilde{x}^2-48}\right)^{1/3}\,, \notag \\
\tilde{x}\,\equiv&\,\,9s_1\,,
\end{align}
and $x_1$ is a solution of $p(x)=0$.


Approaching $x\rightarrow{x}_1$, the metric becomes to be
\begin{equation}
ds_4^2\,\approx\,\left(x_1+s_1\right)^{3/2}x_1^{1/2}\left[\frac{1}{4}ds_{AdS_2}^2+\frac{4\left[d\sigma^2+\mathcal{E}^2(s_1)\sigma^2d\psi^2\right]}{-p'(x_1)}\right]\,,
\end{equation}
where we introduced a new coordinate, $\sigma^2\,=\,x_1-x$. The function, $\mathcal{E}(s_1)$, is given by
\begin{equation} \label{eq13}
\mathcal{E}(s_1)^2\,=\,\frac{p'(x_1)^2}{16\left(p(x_1)+4x_1^2\right)}\,.
\end{equation}
Then, the $x-\psi$ surface is locally an $\mathbb{R}^2/\mathbb{Z}_\ell$ orbifold if we set
\begin{equation} \label{mathcalE3}
\mathcal{E}(s_1)\,=\,\frac{1}{\mathcal{D}k}\,=\,\frac{2\pi}{\Delta\psi{k}}\,,
\end{equation}
where $\Delta\psi$ is the period of the coordinate, $\psi$, and $k=1,2,3,\ldots$. Thus the metric on the surface, $\Sigma(x,\psi)$, has a topology of disk with the origin at $x=x_1$ and the boundary at $x=0$.

We calculate the Euler characteristic of the surface, $\Sigma(x,\psi)$,
\begin{equation}
\chi\left(\Sigma\right)\,=\,\frac{1}{4\pi}\int_\Sigma{R}_\Sigma\text{vol}_\Sigma\,=\,\mathcal{E}\left(s_1\right)\frac{\Delta\psi}{2\pi}\,=\,\mathcal{D}\mathcal{E}\left(s_1\right)\,=\,\frac{1}{k}\,,
\end{equation}
which is a natural result for a disk.

Further analysis of the disk solutions are given in \cite{Suh:2021hef, Couzens:2021rlk} in different coordinate systems.

\section{D4-branes wrapped on a disk} \label{appD}
\renewcommand{\theequation}{D.\arabic{equation}}
\setcounter{equation}{0} 

In this appendix, we consider the $AdS_4\times\text{disk}$ solution of $U(1)^2$-gauged supergravity in six dimensions. As discussed in the introduction, the solution is distinct from the $AdS_4\times\text{disk}$ solution of minimal gauged supergravity in six dimensions, \cite{Suh:2021aik}. The solution and its analysis was already presented in appendix C.1 of \cite{Couzens:2022lvg}. However, in order to have the solution in our conventions, we review in this appendix. Furthermore, we perform the flux quantization and calculate the holographic free energy.

\subsection{The $AdS_4\times\text{disk}$ solutions}

The local form of the $AdS_4\times\text{spindle}$ solution is given by, \cite{Faedo:2021nub},
\begin{align}
ds_6^2\,=&\,\left(y^2h_1h_2\right)^{1/4}\left[ds_{AdS_4}^2+\frac{y^2}{F}dy^2+\frac{F}{h_1h_2}dz^2\right]\,, \notag \\
A_i\,=&\,-\frac{y^3}{h_i}dz\,, \qquad X_i\,=\,\left(y^2h_1h_2\right)^{3/8}h_i^{-1}\,, \notag \\
F(y)\,=&\,m^2h_1h_2-y^4\,, \qquad h_i(y)\,=\,\frac{2g}{3m}y^3+q_i\,,
\end{align}
where $q_i$, $i=1,2$, are constants. The two-form field, $B_{\mu\nu}$, is vanishing.

For the $AdS_4\times\text{disk}$ solutions, we consider the special case of $q_2=0$ and the local form of the solution is given by
\begin{align}
ds_6^2\,=&\,\frac{y^2}{X_1^{2/5}}\left[ds_{AdS_4}^2+\frac{y^2}{F}dy^2+\frac{m^2F}{F+y^4}dz^2\right]\,, \notag \\
X_1\,=&\,\left[\frac{m^2y^6}{F+y^4}\right]^{5/8}\,, \qquad X_2\,=\,\left[\frac{m^2y^6}{F+y^4}\right]^{-3/8}\,, \notag \\
A_1\,=&\,-\frac{m^2y^6}{F+y^4}dz\,, \qquad A_2\,=\,\text{trivial}\,, 
\end{align}
where we employed $2g=3m$ and we have
\begin{equation}
F(y)\,=\,m^2y^3\left(y^3+q_1\right)-y^4\,.
\end{equation}
Thus the solution has a non-trivial scalar field and an $U(1)$ gauge field, effectively. 

The solution can also be given in the form of
\begin{align}
ds_6^2\,=&\,y^{5/4}\left(y^3+q_1\right)^{1/4}\left[ds_{AdS_4}^2+\frac{1}{yh}dy^2+\frac{h}{y^3+q_1}dz^2\right]\,, \notag \\
X_1\,=&\,\left[\frac{y^3}{y^3+q_1}\right]^{5/8}\,, \qquad X_2\,=\,\left[\frac{y^3}{y^3+q_1}\right]^{-3/8}\,, \notag \\
A_1\,=&\,-\frac{y^3}{y^3+q_1}dz\,, \qquad A_2\,=\,\text{trivial}\,, 
\end{align}
where we define
\begin{equation}
h(y)\,=\,m^2\left(y^3+q_1\right)-y\,.
\end{equation}

Now we consider the global completion of the solution, $i.e.$, find the range of the solution where the metric functions are positive definite and the fields are real. We find such solutions when
\begin{equation}
0<y<y_1\,,
\end{equation}
where we have
\begin{align} \label{yone2}
y_1\,=&\,\frac{1}{\sqrt{3}m}\left(\tilde{X}+\frac{1}{\tilde{X}}\right)\,, \notag \\
\tilde{X}\,\equiv&\,\frac{1}{\sqrt{3}}\left(\frac{3}{2}\right)^{1/3}\left(-x+\sqrt{x^2-12}\right)^{1/3}\,, \notag \\
x\,\equiv&\,\,9m^3q_1\,,
\end{align}
and $y_1$ is a solution of $h(y)=0$.

Near $y\rightarrow0$, the $AdS_4$ warp factor vanishes and it is a curvature singularity of the metric,
\begin{equation}
ds_6^2\,\approx\,q_1^{1/4}y^{5/4}\left[ds_{AdS_4}^2+\frac{1}{m^2q_1y}dy^2+m^2dz^2\right]\,.
\end{equation}

Approaching $y\rightarrow{y}_1$, the metric becomes to be
\begin{equation}
ds_6^2\,\approx\,\frac{y_1}{m^{1/2}}\left[ds_{AdS_4}^2+\frac{4\left[d\rho^2+\mathcal{E}^2(q_1)\rho^2dz^2\right]}{y_1\left(-h'(y_1)\right)}\right]\,,
\end{equation}
where we introduced a new coordinate, $\rho^2\,=\,y_1-y$. The function, $\mathcal{E}(q_1)$, is given by
\begin{equation} \label{eq12}
\mathcal{E}(q_1)^2\,=\,\frac{y_1\left(1-3m^2y_1^2\right)^2}{4m\left(y_1^3+q_1\right)}\,.
\end{equation}
Then, the $y-z$ surface is locally an $\mathbb{R}^2/\mathbb{Z}_\ell$ orbifold if we set
\begin{equation} \label{mathcalE2}
\mathcal{E}(q_1)\,=\,\frac{1}{\mathcal{C}\ell}\,=\,\frac{2\pi}{\Delta{z}\ell}\,,
\end{equation}
where $\Delta{z}$ is the period of the coordinate, $z$, and $\ell=1,2,3,\ldots$. Thus the metric on the surface, $\Sigma(y,z)$, has a topology of disk with the origin at $y=y_1$ and the boundary at $y=0$.

We calculate the Euler characteristic of the surface, $\Sigma(y,z)$,
\begin{equation}
\chi\left(\Sigma\right)\,=\,\frac{1}{4\pi}\int_\Sigma{R}_\Sigma\text{vol}_\Sigma\,=\,\frac{2\pi}{4\pi}\frac{y_1^{1/2}\left(2y_1^3-q_1\right)}{\left(y_1^3+q_1\right)^{3/2}}\frac{\Delta{z}}{2\pi}\,=\,\mathcal{E}(q_1)\frac{\Delta{z}}{2\pi}\,=\,\mathcal{C}\mathcal{E}(q_1)\,=\,\frac{1}{\ell}\,,
\end{equation}
which is a natural result for a disk.

\subsection{Uplift to massive type IIA supergravity}

Employing the uplift formula in \cite{Faedo:2021nub, Couzens:2022lvg, Faedo:2022rqx}, with the choice of
\begin{equation}
\mu_0\,=\,\cos\xi\cos\theta\,, \qquad \mu_1\,=\,\sin\xi\,, \qquad \mu_2\,=\,\cos\xi\sin\theta\,,
\end{equation}
and $\xi,\theta\in\left[0,\pi/2\right]$, we uplift the $AdS_4\times\text{disk}$ solutions to massive type IIA supergravity, \cite{Romans:1985tz}. The metric in the string frame, the dilaton, the Romans mass, and the four-form flux are given by, respectively,
\begin{align}
ds_{10}^2\,=&\,\lambda^2\cos^{-1/3}\xi\cos^{-1/3}\theta\tilde{\Delta}^{1/2}y^{1/2}\left[ds_{AdS_4}^2+\frac{1}{yh}dy^2+\frac{h}{y^3+q_1}dz^2\right. \notag \\
&+\left.\frac{1}{g^2y^2}d\xi^2+\frac{1}{g^2\tilde{\Delta}y^2}\Big(\left(y^3+q_1\right)\sin^2\xi{D}\phi_1^2+y^3\cos^2\xi\left(d\theta^2+\sin^2\theta{d}\phi_2\right)\Big)\right]\,, \\
e^\Phi\,=&\,\lambda^2\frac{\left(y^3+q_1\cos^2\xi\right)^{1/4}}{y^{3/4}}\cos^{-5/6}\xi\cos^{-5/6}\theta\,, \\
F_{(0)}\,=&\,\frac{m}{\lambda^3}\,, \\ \label{4ads4}
\lambda^{-1}F_{(4)}\,=&\,-\frac{2}{3g^2}\frac{\left(y^3+q_1\right)\left(5y^3+2q_1\cos^2\xi\right)}{\left(y^3+q_1\cos^2\xi\right)^2}\sin\xi\cos^{7/3}\xi\sin\theta\cos^{1/3}\theta\,d\xi\wedge{d}\theta\wedge{D}\phi_1\wedge{d}\phi_2 \notag \\
&-\frac{3q_1y^2}{g^3\left(y^3+q_1\cos^2\xi\right)^2}\sin^2\xi\cos^{10/3}\xi\sin\theta\cos^{1/3}\theta\,dy\wedge{d}\theta\wedge{D}\phi_1\wedge{d}\phi_2 \notag \\
&-\frac{3q_1y^2}{g^2\left(y^3+q_1\right)\left(y^3+q_1\cos^2\xi\right)}\cos^{10/3}\xi\sin\theta\cos^{1/3}\theta\,dy\wedge{d}z\wedge{d}\theta\wedge{d}\phi_2\,,
\end{align}
where $D\phi_i\equiv{d}\phi_i-A_i$ and $\phi_1,\phi_2\in\left[0,2\pi\right]$. The parameter, $\lambda$, is introduced in the uplift and will be useful for the flux quantizations. We define
\begin{equation}
\tilde{\Delta}\,=\,\left(y^3+q_1\right)\cos^2\xi+y^3\sin^2\xi\,.
\end{equation}
The other fields are trivial.

\subsection{Uplifted metric}

The six-dimensional internal space of the uplifted metric is an $S_z^1\,\times\,S^3$ fibration over the 2d base space, $B_2$, of $(y,\xi)$. The 2d base space is a rectangle of $(y,\xi)$ over $[0,y_1)\,\times\left[0,\frac{\pi}{2}\right]$. It is identical to Figure \ref{fig2} in the main body of the paper. We explain the geometry of the internal space by three regions of the 2d base space, $B_2$.

\begin{itemize}
\item Region I: The side of $\mathsf{P}_1\mathsf{P}_2$.
\item Region II: The sides of $\mathsf{P}_2\mathsf{P}_3$ and $\mathsf{P}_3\mathsf{P}_4$.
\item Region III: The side of $\mathsf{P}_1\mathsf{P}_4$.
\end{itemize}

\noindent {\bf Region I:} On the side of $\xi\,=\,0$, the $\phi_1$ circle, $S^1_{\phi_1}$, shrinks.

\bigskip

\noindent {\bf Region II: Monopole} In order to observe the property of the 2d base, \eqref{2db2}, we fix the gauge of the six-dimensional gauge field to be
\begin{equation}
A_1\,=\,-\frac{y^3}{y^3+q_1}dz-\frac{1}{3}\,.
\end{equation}
We break $D\phi_1^2=\left(d\phi_1-A_1\right)^2$ and complete the square of $dz$ to obtain the metric of
\begin{align}
ds_{10}^2\,=&\,\lambda^2\cos^{-1/3}\xi\cos^{-1/3}\theta\tilde{\Delta}^{1/2}y^{1/2}\left[ds_{AdS_4}^2+\frac{1}{yh}dy^2\right. \notag \\
&+\frac{1}{g^2y^2}d\xi^2+\frac{1}{g^2\tilde{\Delta}y^2}y^3\cos^2\xi\left(d\theta^2+\sin^2\theta{d}\phi_2\right) \notag \\
&+R_z^2\left(dz-gLd\phi_1\right)^2+R_{\phi_1}^2d\phi_1^2\Big]\,,
\end{align}
where we define
\begin{align}
R_z^2\,=&\,\frac{9g^2h_1hy^2\cos^2\xi+\left(h_1\left(h_1-6y^3\right)+9y^5\left(g^2h+y\right)\right)\sin^2\xi}{9g^2y^2h_1\tilde{\Delta}}\,, \notag \\
R_{\phi_1}^2\,=&\,\frac{9h_1h\sin^2\xi}{9g^2h_1hy^2\cos^2\xi+\left(h_1\left(h_1-6y^3\right)+9y^5\left(g^2h+y\right)\right)\sin^2\xi}\,, \notag \\
L\,=&\,\frac{3h_1\left(h_1-3y^2\right)\sin^2\xi}{g\left[9g^2h_1hy^2\cos^2\xi+\left(h_1\left(h_1-6y^3\right)+9y^5\left(g^2h+y\right)\right)\sin^2\xi\right]}\,,
\end{align}
and recall that $h_1=y^3+q_1$.

The function, $L(y,\xi)$, is piecewise constant along the sides of $y=y_1$ and $\xi=\frac{\pi}{2}$ of the 2d base, $B_2$, 
\begin{equation} \label{2db2}
L\left(y,\frac{\pi}{2}\right)\,=\,0\,, \qquad L(y_1,\xi)\,=\,\frac{1}{\mathcal{E}(q_1)}\,=\,\mathcal{C}\ell\,=\,\frac{\Delta{z}}{2\pi}\ell\,.
\end{equation}
The jump at the corner, $(y,\xi)=\left(y_1,\frac{\pi}{2}\right)$, indicates a monopole source for the $D\phi_1$ fibration.

\bigskip

\noindent {\bf Region III: Smeared D4-D8-branes} We introduce a new parametrization of the coordinate, $r\equiv{y}^{3/2}$. In the limit, $r\rightarrow0$, the metric asymptotes to
\begin{align}
ds_{10}^2\,\approx&\,\lambda^2q_1^{1/2}\cos^{2/3}\xi\cos^{-1/3}\theta{r}^{-1/3}\Big[r^{2/3}\left(ds_{AdS_4}^2+m^2dz^2\right) \notag \\
&+\left.\frac{1}{g^2q_1}r^{-2/3}\Big(q_1\left(d\xi^2+\tan^2\xi{D}\phi_1^2\right)+dr^2+r^2\left(d\theta^2+\sin^2\theta{d}\phi_2^2\right)\Big)\right]\,.
\end{align}
The metric implies the smeared D4-D8-brane sources. The D4-D8-branes are 
\begin{itemize}
\item extended along the $AdS_4$ and $z$ directions;
\item localized at the origin of the $\mathbb{R}^3$ parametrized by $S^2$ and $r$, $ds^2_{\mathbb{R}^3}=dr^2+r^2ds^2_{S^2}$;
\item smeared along the $\xi$ and $\phi_1$ directions. 
\end{itemize}
This matches the metric of the smeared D4-D8-branes obtained in the appendix B of \cite{Bah:2017wxp}.

\subsection{Flux quantization}

For the flux quantization, we follow the analysis in \cite{Suh:2021aik}. 

The quantization condition on the Romans mass is given by
\begin{equation}
\left(2\pi\ell_s\right)F_{(0)}\,\equiv\,n_0\,\in\,\mathbb{Z}\,,
\end{equation}
where $\ell_s$ is the string length, $n_0=8-N_f$ and $N_f$ is the number of D8-branes. 

The integral of four-form flux through any four-cycle is an integer,
\begin{equation} \label{romansq2}
\frac{1}{\left(2\pi\ell_s\right)^3}\int_{M_4}F_{(4)}\,\in\,\mathbb{Z}\,.
\end{equation}
First, from the flux through the four-cylce of $\xi\theta\phi_1\phi_2$, we obtain
\begin{equation} \label{four1q2}
\frac{1}{\left(2\pi\ell_s\right)^3}\int_{\xi\theta\phi_1\phi_2}{F}_{(4)}\,=\,\frac{3\lambda}{8\pi\ell_s^3g^3}\,\equiv\,N\,,
\end{equation}
where $N\in\mathbb{N}$ is the number of D4-branes wrapping the disk, $\Sigma$. From \eqref{romansq2} and \eqref{four1q2}, we find, \cite{Faedo:2021nub},
\begin{equation} \label{fluxglam2}
g^8\,=\,\frac{1}{\left(2\pi\ell_s\right)^8}\frac{18\pi^6}{N^3n_0}\,, \qquad \lambda^8\,=\,\frac{8\pi^2}{9Nn_0^3}\,.
\end{equation}

Second, from the flux through the four cycle of $yz\theta\phi_2$ including the disk, $\Sigma(y,z)$, at $\xi=0$, we find
\begin{equation} \label{qK2}
\frac{1}{\left(2\pi\ell_s\right)^3}\int_{yz\theta\phi_2}{F}_{(4)}\,=\,-\frac{1}{\left(2\pi\ell_s\right)^3}\frac{3\lambda}{4g^2}\frac{y_1^3}{y_1^3+q_1}2\pi\Delta{z}\,=\,-\frac{N}{\ell}\frac{1}{\mathcal{E}(q_1)}\frac{gy_1^3}{y_1^3+q_1}\,\equiv\,-K\,,
\end{equation}
where $K\in\mathbb{Z}$ is another integer. By eliminating $\mathcal{E}(q_1)$ from the constraints, \eqref{mathcalE2} and \eqref{qK2}, and with the expression of $y_1(q_1)$ in \eqref{yone2}, we obtain
\begin{equation} \label{q1one2}
q_1\,=\,\frac{K^{1/2}\left(g\,\mathcal{C}N-K\right)}{m^3\left(g\,\mathcal{C}N\right)^{3/2}}\,,
\end{equation}
and we also find
\begin{equation} \label{yoneyone2}
y_1\,=\,\frac{1}{m}\sqrt{\frac{K}{g\,\mathcal{C}N}}\,.
\end{equation}
Then, by plugging $y_1(q_1)$, \eqref{yoneyone2}, in \eqref{mathcalE2} with the expression of $\mathcal{E}(q_1)$ in \eqref{eq12}, we also find another expression for $q_1$,
\begin{equation} \label{q1two2}
q_1\,=\,\frac{K^{1/2}\left(9m^2\mathcal{C}K^2\ell^2-4gNK-6gm^2\mathcal{C}^2NK\ell^2+gm^2\mathcal{C}^3N^2\ell^2\right)}{4g^{5/2}m^3\mathcal{C}^{3/2}N^{5/2}}\,.
\end{equation}
Finally, identifying \eqref{q1one2} and \eqref{q1two2}, we can solve for $\mathcal{C}$ and then for $q_1$ in terms of the quantum numbers, $N$, $K$, and $\ell$,
\begin{equation}
\mathcal{C}\,=\,\frac{3\left(N+K\ell\right)}{gN\ell}\,, \qquad q_1\,=\,\frac{1}{3\sqrt{3}m^3}\frac{\sqrt{K\ell}\left(3N+2K\ell\right)}{\left(N+K\ell\right)^{3/2}}\,.
\end{equation}

\subsection{Holographic free energy}

Now we calculate the holographic free energy of dual 3d SCFTs. For the metric in the string frame,
\begin{equation}
ds_{10}^2\,=\,e^{2\mathcal{A}}\left(ds_{AdS_4}^2+ds_{M_6}^2\right)\,,
\end{equation}
the holographic free energy is given by, $e.g.$, \cite{Faedo:2021nub}
\begin{equation}
\mathcal{F}\,=\,\frac{16\pi^3}{\left(2\pi\ell_s\right)^8}\int_{M_6}e^{8\mathcal{A}-2\Phi}\text{vol}_{M_6}\,.
\end{equation}
From above, we obtain
\begin{equation} \label{hfree}
\mathcal{F}\,=\,\frac{16\pi^3}{\left(2\pi\ell_s\right)^8}\frac{3\pi^2\lambda^4}{10g^4}y_1^3\Delta{z}\,=\,\frac{2\sqrt{6}\pi}{5\sqrt{8-N_f}}\sqrt{\frac{N^3K^3\ell}{N+K\ell}}\,.
\end{equation}
If we have $K\sim{N}$, the free energy scales as $\mathcal{F}\sim{N}^{5/2}$ as the free energy of 5d SCFTs.

\bibliographystyle{JHEP}
\bibliography{20241008}

\providecommand{\href}[2]{#2}\begingroup\raggedright\begin{thebibliography}{10}

\bibitem{Ferrero:2020laf}
P.~Ferrero, J.~P. Gauntlett, J.~M. P\'erez Ipi\~na, D.~Martelli and J.~Sparks,
  \emph{{D3-Branes Wrapped on a Spindle}},
  \href{http://dx.doi.org/10.1103/PhysRevLett.126.111601}{\emph{Phys. Rev.
  Lett.} {\bf 126} (2021) 111601},
  [\href{https://arxiv.org/abs/2011.10579}{{\tt 2011.10579}}].

\bibitem{Ferrero:2020twa}
P.~Ferrero, J.~P. Gauntlett, J.~M. P\'erez Ipi\~na, D.~Martelli and J.~Sparks,
  \emph{{Accelerating black holes and spinning spindles}},
  \href{http://dx.doi.org/10.1103/PhysRevD.104.046007}{\emph{Phys. Rev. D} {\bf
  104} (2021) 046007}, [\href{https://arxiv.org/abs/2012.08530}{{\tt
  2012.08530}}].

\bibitem{Bah:2021mzw}
I.~Bah, F.~Bonetti, R.~Minasian and E.~Nardoni, \emph{{Holographic Duals of
  Argyres-Douglas Theories}},
  \href{http://dx.doi.org/10.1103/PhysRevLett.127.211601}{\emph{Phys. Rev.
  Lett.} {\bf 127} (2021) 211601},
  [\href{https://arxiv.org/abs/2105.11567}{{\tt 2105.11567}}].

\bibitem{Bah:2021hei}
I.~Bah, F.~Bonetti, R.~Minasian and E.~Nardoni, \emph{{M5-brane sources,
  holography, and Argyres-Douglas theories}},
  \href{http://dx.doi.org/10.1007/JHEP11(2021)140}{\emph{JHEP} {\bf 11} (2021)
  140}, [\href{https://arxiv.org/abs/2106.01322}{{\tt 2106.01322}}].

\bibitem{Suh:2021hef}
M.~Suh, \emph{{M2-branes wrapped on a topological disk}},
  \href{http://dx.doi.org/10.1007/JHEP09(2022)048}{\emph{JHEP} {\bf 09} (2022)
  048}, [\href{https://arxiv.org/abs/2109.13278}{{\tt 2109.13278}}].

\bibitem{Couzens:2021rlk}
C.~Couzens, K.~Stemerdink and D.~van~de Heisteeg, \emph{{M2-branes on discs and
  multi-charged spindles}},
  \href{http://dx.doi.org/10.1007/JHEP04(2022)107}{\emph{JHEP} {\bf 04} (2022)
  107}, [\href{https://arxiv.org/abs/2110.00571}{{\tt 2110.00571}}].

\bibitem{Couzens:2021tnv}
C.~Couzens, N.~T. Macpherson and A.~Passias, \emph{{$ \mathcal{N} $ = (2, 2)
  AdS$_{3}$ from D3-branes wrapped on Riemann surfaces}},
  \href{http://dx.doi.org/10.1007/JHEP02(2022)189}{\emph{JHEP} {\bf 02} (2022)
  189}, [\href{https://arxiv.org/abs/2107.13562}{{\tt 2107.13562}}].

\bibitem{Suh:2021ifj}
M.~Suh, \emph{{D3-branes and M5-branes wrapped on a topological disc}},
  \href{http://dx.doi.org/10.1007/JHEP03(2022)043}{\emph{JHEP} {\bf 03} (2022)
  043}, [\href{https://arxiv.org/abs/2108.01105}{{\tt 2108.01105}}].

\bibitem{Boisvert:2024jrl}
M.~Boisvert and P.~Ferrero, \emph{{A story of non-conformal branes: spindles,
  disks, circles and black holes}},
  \href{http://dx.doi.org/10.1007/JHEP06(2024)013}{\emph{JHEP} {\bf 06} (2024)
  013}, [\href{https://arxiv.org/abs/2403.03989}{{\tt 2403.03989}}].

\bibitem{Suh:2021aik}
M.~Suh, \emph{{D4-branes wrapped on a topological disk}},
  \href{http://dx.doi.org/10.1007/JHEP06(2023)008}{\emph{JHEP} {\bf 06} (2023)
  008}, [\href{https://arxiv.org/abs/2108.08326}{{\tt 2108.08326}}].

\bibitem{Couzens:2022lvg}
C.~Couzens, H.~Kim, N.~Kim, Y.~Lee and M.~Suh, \emph{{D4-branes wrapped on
  four-dimensional orbifolds through consistent truncation}},
  \href{http://dx.doi.org/10.1007/JHEP02(2023)025}{\emph{JHEP} {\bf 02} (2023)
  025}, [\href{https://arxiv.org/abs/2210.15695}{{\tt 2210.15695}}].

\bibitem{Maldacena:1997re}
J.~M. Maldacena, \emph{{The Large N limit of superconformal field theories and
  supergravity}}, \href{http://dx.doi.org/10.1023/A:1026654312961}{\emph{Adv.
  Theor. Math. Phys.} {\bf 2} (1998) 231--252},
  [\href{https://arxiv.org/abs/hep-th/9711200}{{\tt hep-th/9711200}}].

\bibitem{Argyres:1995jj}
P.~C. Argyres and M.~R. Douglas, \emph{{New phenomena in SU(3) supersymmetric
  gauge theory}},
  \href{http://dx.doi.org/10.1016/0550-3213(95)00281-V}{\emph{Nucl. Phys. B}
  {\bf 448} (1995) 93--126}, [\href{https://arxiv.org/abs/hep-th/9505062}{{\tt
  hep-th/9505062}}].

\bibitem{Karndumri:2022wpu}
P.~Karndumri and P.~Nuchino, \emph{{Five-branes wrapped on topological disks
  from 7D N=2 gauged supergravity}},
  \href{http://dx.doi.org/10.1103/PhysRevD.105.066010}{\emph{Phys. Rev. D} {\bf
  105} (2022) 066010}, [\href{https://arxiv.org/abs/2201.05037}{{\tt
  2201.05037}}].

\bibitem{Couzens:2022yjl}
C.~Couzens, H.~Kim, N.~Kim and Y.~Lee, \emph{{Holographic duals of M5-branes on
  an irregularly punctured sphere}},
  \href{http://dx.doi.org/10.1007/JHEP07(2022)102}{\emph{JHEP} {\bf 07} (2022)
  102}, [\href{https://arxiv.org/abs/2204.13537}{{\tt 2204.13537}}].

\bibitem{Bah:2022yjf}
I.~Bah, F.~Bonetti, E.~Nardoni and T.~Waddleton, \emph{{Aspects of irregular
  punctures via holography}},
  \href{http://dx.doi.org/10.1007/JHEP11(2022)131}{\emph{JHEP} {\bf 11} (2022)
  131}, [\href{https://arxiv.org/abs/2207.10094}{{\tt 2207.10094}}].

\bibitem{Bomans:2023ouw}
P.~Bomans, C.~Couzens, Y.~Lee and S.~Ning, \emph{{Symmetry breaking and
  consistent truncations from M5-branes wrapping a disc}},
  \href{http://dx.doi.org/10.1007/JHEP01(2024)088}{\emph{JHEP} {\bf 01} (2024)
  088}, [\href{https://arxiv.org/abs/2308.08616}{{\tt 2308.08616}}].

\bibitem{Couzens:2023kyf}
C.~Couzens, M.~J. Kang, C.~Lawrie and Y.~Lee, \emph{{Holographic duals of
  Higgsed $\mathcal{D}_p^b(BCD)$}},
  \href{https://arxiv.org/abs/2312.12503}{{\tt 2312.12503}}.

\bibitem{Gauntlett:2001jj}
J.~P. Gauntlett and N.~Kim, \emph{{M five-branes wrapped on supersymmetric
  cycles. 2.}}, \href{http://dx.doi.org/10.1103/PhysRevD.65.086003}{\emph{Phys.
  Rev. D} {\bf 65} (2002) 086003},
  [\href{https://arxiv.org/abs/hep-th/0109039}{{\tt hep-th/0109039}}].

\bibitem{Suh:2018tul}
M.~Suh, \emph{{Supersymmetric AdS$_{6}$ black holes from F(4) gauged
  supergravity}}, \href{http://dx.doi.org/10.1007/JHEP01(2019)035}{\emph{JHEP}
  {\bf 01} (2019) 035}, [\href{https://arxiv.org/abs/1809.03517}{{\tt
  1809.03517}}].

\bibitem{Hosseini:2018usu}
S.~M. Hosseini, K.~Hristov, A.~Passias and A.~Zaffaroni, \emph{{6D attractors
  and black hole microstates}},
  \href{http://dx.doi.org/10.1007/JHEP12(2018)001}{\emph{JHEP} {\bf 12} (2018)
  001}, [\href{https://arxiv.org/abs/1809.10685}{{\tt 1809.10685}}].

\bibitem{Suh:2018szn}
M.~Suh, \emph{{Supersymmetric $AdS_6$ black holes from matter coupled $F(4)$
  gauged supergravity}},
  \href{http://dx.doi.org/10.1007/JHEP02(2019)108}{\emph{JHEP} {\bf 02} (2019)
  108}, [\href{https://arxiv.org/abs/1810.00675}{{\tt 1810.00675}}].

\bibitem{Kim:2019fsg}
N.~Kim and M.~Shim, \emph{{Wrapped Brane Solutions in Romans $F(4)$ Gauged
  Supergravity}},
  \href{http://dx.doi.org/10.1016/j.nuclphysb.2019.114882}{\emph{Nucl. Phys. B}
  {\bf 951} (2020) 114882}, [\href{https://arxiv.org/abs/1909.01534}{{\tt
  1909.01534}}].

\bibitem{Boido:2021szx}
A.~Boido, J.~M.~P. Ipi\~na and J.~Sparks, \emph{{Twisted D3-brane and M5-brane
  compactifications from multi-charge spindles}},
  \href{http://dx.doi.org/10.1007/JHEP07(2021)222}{\emph{JHEP} {\bf 07} (2021)
  222}, [\href{https://arxiv.org/abs/2104.13287}{{\tt 2104.13287}}].

\bibitem{Suh:2022olh}
M.~Suh, \emph{{M5-branes and D4-branes wrapped on a direct product of spindle
  and Riemann surface}},
  \href{http://dx.doi.org/10.1007/JHEP02(2024)205}{\emph{JHEP} {\bf 02} (2024)
  205}, [\href{https://arxiv.org/abs/2207.00034}{{\tt 2207.00034}}].

\bibitem{Giri:2021xta}
S.~Giri, \emph{{Black holes with spindles at the horizon}},
  \href{http://dx.doi.org/10.1007/JHEP06(2022)145}{\emph{JHEP} {\bf 06} (2022)
  145}, [\href{https://arxiv.org/abs/2112.04431}{{\tt 2112.04431}}].

\bibitem{Faedo:2021nub}
F.~Faedo and D.~Martelli, \emph{{D4-branes wrapped on a spindle}},
  \href{http://dx.doi.org/10.1007/JHEP02(2022)101}{\emph{JHEP} {\bf 02} (2022)
  101}, [\href{https://arxiv.org/abs/2111.13660}{{\tt 2111.13660}}].

\bibitem{Cheung:2022ilc}
K.~C.~M. Cheung, J.~H.~T. Fry, J.~P. Gauntlett and J.~Sparks, \emph{{M5-branes
  wrapped on four-dimensional orbifolds}},
  \href{http://dx.doi.org/10.1007/JHEP08(2022)082}{\emph{JHEP} {\bf 08} (2022)
  082}, [\href{https://arxiv.org/abs/2204.02990}{{\tt 2204.02990}}].

\bibitem{Faedo:2022rqx}
F.~Faedo, A.~Fontanarossa and D.~Martelli, \emph{{Branes wrapped on orbifolds
  and their gravitational blocks}},
  \href{http://dx.doi.org/10.1007/s11005-023-01671-1}{\emph{Lett. Math. Phys.}
  {\bf 113} (2023) 51}, [\href{https://arxiv.org/abs/2210.16128}{{\tt
  2210.16128}}].

\bibitem{Faedo:2024upq}
F.~Faedo, A.~Fontanarossa and D.~Martelli, \emph{{Branes wrapped on
  quadrilaterals}},  \href{https://arxiv.org/abs/2402.08724}{{\tt 2402.08724}}.

\bibitem{Liu:1999ai}
J.~T. Liu and R.~Minasian, \emph{{Black holes and membranes in AdS(7)}},
  \href{http://dx.doi.org/10.1016/S0370-2693(99)00500-6}{\emph{Phys. Lett. B}
  {\bf 457} (1999) 39--46}, [\href{https://arxiv.org/abs/hep-th/9903269}{{\tt
  hep-th/9903269}}].

\bibitem{Pernici:1984xx}
M.~Pernici, K.~Pilch and P.~van Nieuwenhuizen, \emph{{Gauged Maximally Extended
  Supergravity in Seven-dimensions}},
  \href{http://dx.doi.org/10.1016/0370-2693(84)90813-X}{\emph{Phys. Lett. B}
  {\bf 143} (1984) 103--107}.

\bibitem{Cvetic:2000ah}
M.~Cvetic, H.~Lu, C.~N. Pope, A.~Sadrzadeh and T.~A. Tran, \emph{{S**3 and S**4
  reductions of type IIA supergravity}},
  \href{http://dx.doi.org/10.1016/S0550-3213(00)00466-1}{\emph{Nucl. Phys. B}
  {\bf 590} (2000) 233--251}, [\href{https://arxiv.org/abs/hep-th/0005137}{{\tt
  hep-th/0005137}}].

\bibitem{Cremmer:1978km}
E.~Cremmer, B.~Julia and J.~Scherk, \emph{{Supergravity Theory in 11
  Dimensions}},
  \href{http://dx.doi.org/10.1016/0370-2693(78)90894-8}{\emph{Phys. Lett. B}
  {\bf 76} (1978) 409--412}.

\bibitem{Ferrero:2021etw}
P.~Ferrero, J.~P. Gauntlett and J.~Sparks, \emph{{Supersymmetric spindles}},
  \href{http://dx.doi.org/10.1007/JHEP01(2022)102}{\emph{JHEP} {\bf 01} (2022)
  102}, [\href{https://arxiv.org/abs/2112.01543}{{\tt 2112.01543}}].

\bibitem{Romans:1985tw}
L.~J. Romans, \emph{{The F(4) Gauged Supergravity in Six-dimensions}},
  \href{http://dx.doi.org/10.1016/0550-3213(86)90517-1}{\emph{Nucl. Phys. B}
  {\bf 269} (1986) 691}.

\bibitem{Andrianopoli:2001rs}
L.~Andrianopoli, R.~D'Auria and S.~Vaula, \emph{{Matter coupled F(4) gauged
  supergravity Lagrangian}},
  \href{http://dx.doi.org/10.1088/1126-6708/2001/05/065}{\emph{JHEP} {\bf 05}
  (2001) 065}, [\href{https://arxiv.org/abs/hep-th/0104155}{{\tt
  hep-th/0104155}}].

\bibitem{Karndumri:2015eta}
P.~Karndumri, \emph{{Twisted compactification of N = 2 5D SCFTs to three and
  two dimensions from F(4) gauged supergravity}},
  \href{http://dx.doi.org/10.1007/JHEP09(2015)034}{\emph{JHEP} {\bf 09} (2015)
  034}, [\href{https://arxiv.org/abs/1507.01515}{{\tt 1507.01515}}].

\bibitem{Romans:1985tz}
L.~J. Romans, \emph{{Massive N=2a Supergravity in Ten-Dimensions}},
  \href{http://dx.doi.org/10.1016/0370-2693(86)90375-8}{\emph{Phys. Lett. B}
  {\bf 169} (1986) 374}.

\bibitem{Bah:2017wxp}
I.~Bah, A.~Passias and A.~Tomasiello, \emph{{AdS$_{5}$ compactifications with
  punctures in massive IIA supergravity}},
  \href{http://dx.doi.org/10.1007/JHEP11(2017)050}{\emph{JHEP} {\bf 11} (2017)
  050}, [\href{https://arxiv.org/abs/1704.07389}{{\tt 1704.07389}}].

\bibitem{Couzens:2021cpk}
C.~Couzens, \emph{{A tale of (M)2 twists}},
  \href{http://dx.doi.org/10.1007/JHEP03(2022)078}{\emph{JHEP} {\bf 03} (2022)
  078}, [\href{https://arxiv.org/abs/2112.04462}{{\tt 2112.04462}}].

\bibitem{Hosseini:2018uzp}
S.~M. Hosseini, I.~Yaakov and A.~Zaffaroni, \emph{{Topologically twisted
  indices in five dimensions and holography}},
  \href{http://dx.doi.org/10.1007/JHEP11(2018)119}{\emph{JHEP} {\bf 11} (2018)
  119}, [\href{https://arxiv.org/abs/1808.06626}{{\tt 1808.06626}}].

\bibitem{Crichigno:2018adf}
P.~M. Crichigno, D.~Jain and B.~Willett, \emph{{5d Partition Functions with A
  Twist}}, \href{http://dx.doi.org/10.1007/JHEP11(2018)058}{\emph{JHEP} {\bf
  11} (2018) 058}, [\href{https://arxiv.org/abs/1808.06744}{{\tt 1808.06744}}].

\bibitem{Brandhuber:1999np}
A.~Brandhuber and Y.~Oz, \emph{{The D-4 - D-8 brane system and five-dimensional
  fixed points}},
  \href{http://dx.doi.org/10.1016/S0370-2693(99)00763-7}{\emph{Phys. Lett. B}
  {\bf 460} (1999) 307--312}, [\href{https://arxiv.org/abs/hep-th/9905148}{{\tt
  hep-th/9905148}}].

\bibitem{Seiberg:1996bd}
N.~Seiberg, \emph{{Five-dimensional SUSY field theories, nontrivial fixed
  points and string dynamics}},
  \href{http://dx.doi.org/10.1016/S0370-2693(96)01215-4}{\emph{Phys. Lett. B}
  {\bf 388} (1996) 753--760}, [\href{https://arxiv.org/abs/hep-th/9608111}{{\tt
  hep-th/9608111}}].

\end{thebibliography}\endgroup

\end{document}